\documentclass[12pt]{iopart}
\usepackage{graphicx}
\begin{document}
\title[Collimator ef\mbox{}fects in proton planning]{Collimator ef\mbox{}fects in proton planning}

\author{Evangelos Matsinos}

\address{Varian Medical Systems Imaging Laboratory GmbH, T\"{a}fernstrasse 7, CH-5405 Baden-D\"{a}ttwil, Switzerland}
\ead{evangelos.matsinos@varian.com~\footnote{E-mail address after December 31, 2008: evangelos[dot]matsinos[at]sunrise[dot]ch}}

\begin{abstract}
The present paper pertains to corrections which are due to the presence of beam-limiting and beam-shaping devices in proton planning. 
Two types of corrections are considered: those originating from the nonzero thickness of such devices (geometrical ef\mbox{}fects) 
and those relating to the scattering of beam particles of\mbox{}f their material. The application of these two types of corrections 
is greatly facilitated by decomposing their ef\mbox{}fects on the f\mbox{}luence into easily-calculable contributions. To expedite 
the derivation of the scattering corrections, a two-step process has been introduced into a commercial product which is widely used 
in planning (Treatment Planning System Eclipse$^{\rm (TM)}$). The f\mbox{}irst step occurs at the beam-conf\mbox{}iguration phase and 
comprises the analysis of half-block f\mbox{}luence measurements and the extraction of the one parameter of the model which is used in 
the description of the beamline characteristics; subsequently, a number of Monte-Carlo runs yield the parameters of a convenient 
parameterisation of the relevant f\mbox{}luence contributions. The second step involves (at planning time) the reconstruction of the 
parameters (which are used in the description of the scattering contributions) via simple interpolations, performed on the results 
obtained during the beam-conf\mbox{}iguration phase. Given the lack of dedicated data, the validation of the method is currently based 
on the reproduction of the parts of the half-block f\mbox{}luence measurements (i.e., of the data used as input during the beam 
conf\mbox{}iguration) which had been removed from the database to suppress unwanted (block-scattering) contributions; it is convincingly 
demonstrated that the inclusion of the scattering ef\mbox{}fects leads to substantial improvement in the reproduction of the experimental 
data. The contributions from the block-thickness and block-scattering ef\mbox{}fects (to the f\mbox{}luence) are presented separately in 
the case of a simple water phantom; in this example, the maximal contribution of the two block-relating ef\mbox{}fects amounts to a few 
percent of the prescribed dose.
\end{abstract}

\section{\label{sec:Introduction}Introduction}

To shape the beam so that it matches the characteristics of the specif\mbox{}ic treatment in radiation therapy (thus achieving the 
delivery of the prescribed dose to the target (tumour) and maximal protection of the surrounding healthy tissue and vital 
organs), beam-limiting and beam-shaping devices (BL/BSDs) are routinely used. Generally speaking, the beam is f\mbox{}irst restricted 
(in size) by the primary collimator, a beam-limiting device giving it a rectangular shape. The beam may subsequently encounter 
the multi-leaf collimator (MLC), which may be static or dynamic (i.e., undergoing software-controlled motion during the treatment 
session). More frequently than not, the desirable beam shaping is achieved by inserting a metallic piece (with the appropriate 
aperture and thickness) into the beamline, directly in front of the patient; this last beam-shaping device is called a patient 
collimator or simply a block. Being positioned close to the patient, the block achieves ef\mbox{}f\mbox{}icient fall-of\mbox{}f of 
the dose (sharp penumbra) outside the target area. (The simultaneous use of MLC and block is not common.)

To provide ef\mbox{}f\mbox{}icient attenuation of the beam outside the irradiated volume, the BL/BSDs are made of high-$Z$ materials. 
The primary collimator and the MLC are f\mbox{}ixed parts of the treatment machine, whereas the block not only depends on the 
particular patient, but also on the direction from which the target is irradiated. Therefore, there may be several blocks in one 
treatment plan, not necessarily corresponding to the same thickness or (perhaps) material.

The presence of BL/BSDs in treatment plans induces three types of physical ef\mbox{}fects:
\begin{itemize}
\item[] a) Conf\mbox{}inement of the beam to the area corresponding to full transmission (i.e., to the aperture of the device).
\item[] b) Ef\mbox{}fects associated with the nonzero thickness of the device (geometrical ef\mbox{}fects).
\item[] c) Ef\mbox{}fects relating to the scattering of the beam of\mbox{}f the material of the device.
\end{itemize}
Type-(a) ef\mbox{}fects (direct blocking of the beam) are dominant and have always been taken into account. The standard way to do this 
is by reducing the BL/BSD into a two-dimensional (2D) object (i.e., by disregarding its thickness) and assuming no transmission 
of the beam outside its aperture. Type-(b) and type-(c) ef\mbox{}fects induce corrections which, albeit at a few-percent level of the 
prescribed dose, may represent a sizable fraction of the \emph{local} dose; due to their complexity and to time restrictions during 
the planning phase, these corrections have (so far) been omitted in clinical applications.

The subject of the slit scattering in beam collimation was f\mbox{}irst addressed by Courant (1951). Courant extracted analytical 
solutions for the ef\mbox{}fective increase in the slit width (attributable to scattering) by solving the dif\mbox{}fusion equation inside 
the collimator. To fulf\mbox{}ill the boundary conditions, he introduced the negative-image technique, which was later criticised 
(e.g., by Burge and Smith (1962)). Despite the debatable usefulness of the practical use of Courant's work, that article set forth 
def\mbox{}initions which were employed in future research; for example, by categorising the scattered particles as:
\begin{itemize}
\item those impinging upon the upstream face of the collimator and emerging from its inner surface (bore),
\item those entering the bore and scattering out of it, and
\item those entering the bore and leaving the downstream face of the collimator.
\end{itemize}
In the present paper, Courant's type-1 particles will correspond to our `outer tracks' (OTs), type-2 particles to our `bore-scattered 
inner tracks' (BSITs), and type-3 particles to our `going-through inner tracks' (GTITs), see F\mbox{}ig.~\ref{fig:Model1}. The tracks which 
do not hit the block will comprise the `pristine' beam.

Aiming at the determination of the optimal material for proton-beam collimation and intending to provide data for experiments 
at the linear accelerator of the National Institute for Research in Nuclear Science (NIRNS) at Harwell, Burge and Smith (1962) 
re-addressed the slit-scattering problem and obtained a solution via the numerical integration of the dif\mbox{}fusion equation. 
Burge and Smith reported considerable dif\mbox{}ferences to Courant's results. In the last section of their article, the authors 
discussed alternative approaches to the slit-scattering problem.

A Monte-Carlo (MC) method, as a means to study the collimator-scattering ef\mbox{}fects, was introduced by Paganetti (1998). Using 
the GEANT code to simulate a proton beamline at the Hahn-Meitner Institut (HMI) in Berlin, Paganetti introduced simple parameterisations 
to account for the changes in the energy and angular divergence of the beam as it traverses the various beamline elements. Judging from 
F\mbox{}ig.~6 of that paper, one expects the scattering ef\mbox{}fects to be around the $1 \%$ level. In the last section of his article, 
Paganetti correctly predicted that `Monte-Carlo methods will become important for providing proton phase-space distributions for input to 
treatment-planning routines, though the calculation of the target dose will still be done analytically.' Our strategy is similar to that 
of Paganetti: the dose, delivered by the pristine beam, will be corrected for scattering ef\mbox{}fects on the basis of results obtained 
via MC runs prior to planning (actually, when the particular proton-treatment machine is conf\mbox{}igured).

Block scattering was also investigated, along the general lines of Paganetti's paper, in the work of van Luijk \etal (2001). The 
(same version of the) GEANT code was used, to simulate a proton beamline at Kernfysisch Versneller Instituut (KVI) in Groningen, 
and the characteristics of the scattered protons were studied. To validate their approach, the authors obtained dose measurements 
for several f\mbox{}ield sizes and at several distances from the block. Unlike other works, the ef\mbox{}fect of scattering in air 
was also included in that study, and turned out to be more signif\mbox{}icant than previously thought (its contribution to the angular 
divergence of the beam exceeded $1$ mrad per $1$ m of air). One of the interesting conclusions of that paper was that the penumbra 
of the dose distribution is mostly accounted for by the lateral spread of the undisturbed beam; that conclusion somewhat allayed 
former fears that the extraction of the ef\mbox{}fective-source size ($\sigma_s$) and of the ef\mbox{}fective source-axis distance 
(SAD), both obtained from measurements conducted in the presence of a block, might be seriously af\mbox{}fected by block-scattering 
contributions. F\mbox{}inally, the paper addressed the asymmetries induced by a misalignment of the block, concluding that they 
might be sizable.

In their recent work, Kimstrand \etal (2008) put the emphasis on the scattering of\mbox{}f the multi-leaf collimator (MLC). In their 
approach, they also obtain the values of the parameters (involved in their corrections) by using a MC GEANT-based method. In their 
Section `Discussion and Conclusions', the authors, setting forth the future perspectives, mention that `a challenge \dots is to 
implement collimator scatter for a pencil beam kernel dose calculation engine.' We agree on this being `a challenge'. The authors 
then advance to pre-empt that `\dots the methods presented \dots can straightforwardly be applied to arbitrary shaped collimators 
of dif\mbox{}ferent materials, such as moulded patient-specif\mbox{}ic collimators used in passively scattered proton beams.' One 
should at least remark that their paper does not contain adequate information supporting the thesis that the proposed approach is 
of practical use in a domain where the execution time is a important factor.

\section{\label{sec:Method}Method}

In the present paper, the standard ICRU (1987) coordinate system will be used; in beam's eye view, the $x$ axis points to the 
right, $y$ upward, and $z$ toward the source. The origin of the coordinate system is the isocentre.

\subsection{\label{sec:Miniblocks}Miniblocks}

Assuming vanishing transmission outside the aperture of the BL/BSD, all relevant contributions to the f\mbox{}luence involve 2D 
integrals over its area. The evaluation of such integrals is time-consuming; at a time when serious ef\mbox{}forts are made to 
reduce the overall time allotted to each patient, the generous allocation of resources in the evaluation and application of 
\emph{corrections} to the primary dose is unacceptable. To expedite the extraction of these corrections, one must f\mbox{}ind a fast 
way to decompose the ef\mbox{}fects of a 2D object (e.g., of the aperture of the BL/BSD) into one-dimensional, easily-calculable, 
easily-applicable contributions.

\begin{figure}
\begin{center}
\includegraphics [width=15.5cm] {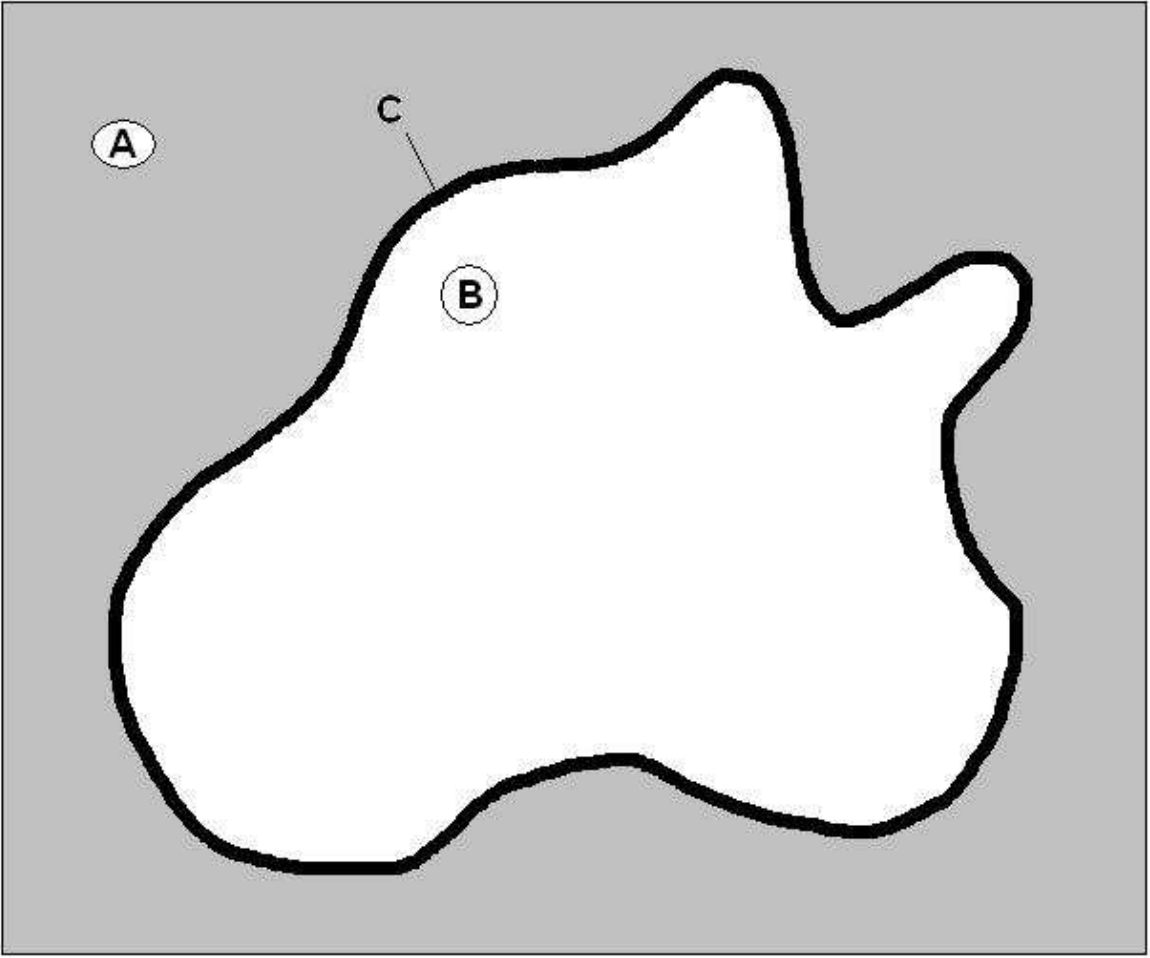}
\caption{\label{fig:Miniblock1}Example of a BL/BSD (reduced to two dimensions) in beam's eye view.}
\end{center}
\end{figure}

An example of one such 2D object is shown in F\mbox{}ig.~\ref{fig:Miniblock1}; the area B represents the aperture (corresponding to full 
transmission of the beam) and is separated from the area A, representing the (high-$Z$) material, via the contour C. Full attenuation of 
the beam is assumed in the area A. The contour C and the `outer' contour of the area A (which is not shown) may have any arbitrary 
shape (e.g., rectangular, circular, etc.); the only requirement is that the contour C be contained within the contour of the area A.

\begin{figure}
\begin{center}
\includegraphics [width=15.5cm] {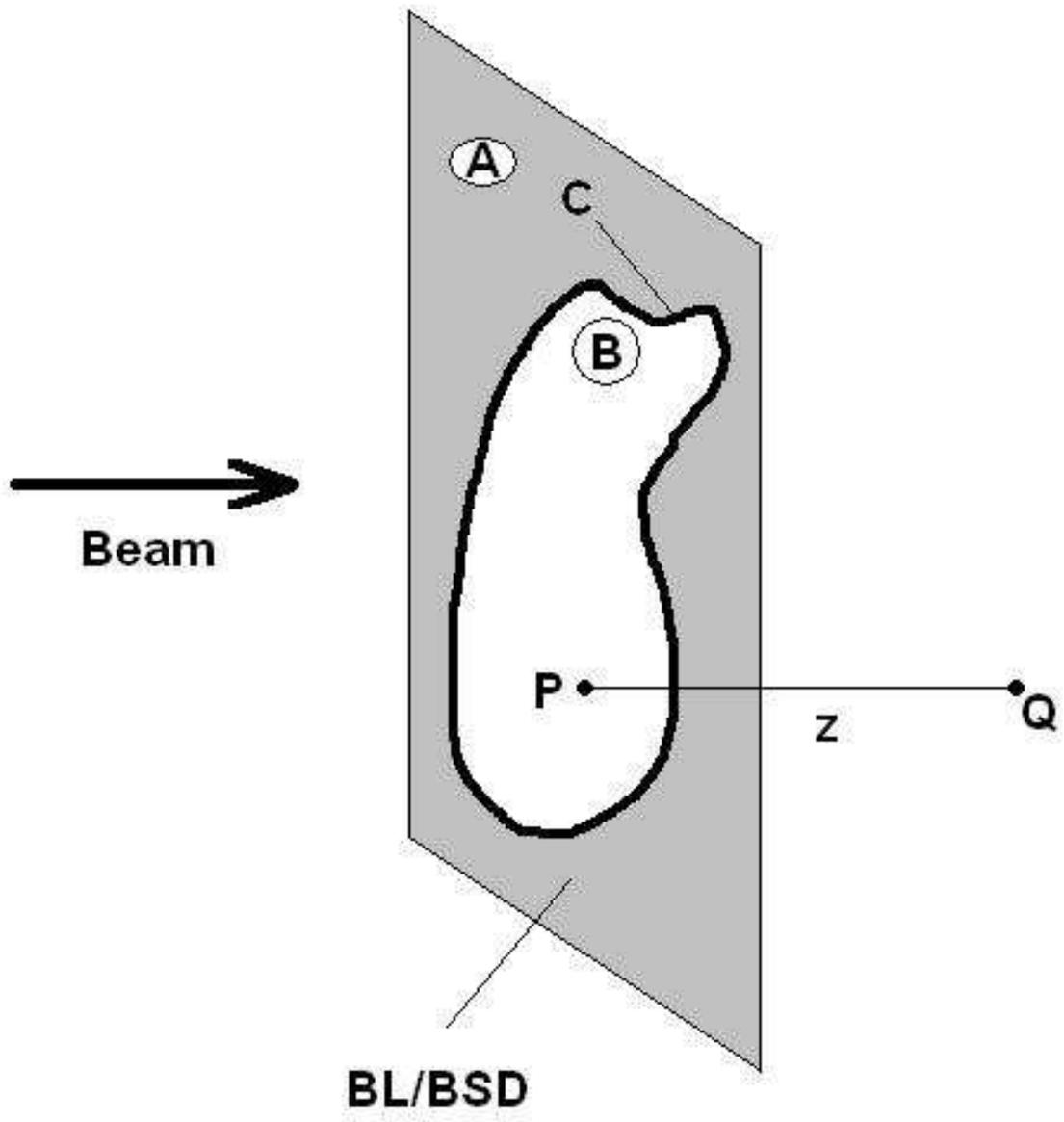}
\caption{\label{fig:Miniblock2}The ef\mbox{}fects of the BL/BSD have to be evaluated at the point Q, which is projected to the point P on 
the BL/BSD plane. In this f\mbox{}igure, the point P lies within the area B.}
\end{center}
\end{figure}

Let us assume that the aim is to derive the inf\mbox{}luence of the BL/BSD at one point Q (F\mbox{}ig.~\ref{fig:Miniblock2}); the point Q is 
projected to the point P, on the BL/BSD plane. The point P may be contained in the interior or the exterior of the area B.

\subsubsection{\label{sec:Within}The point P lies within the area B}

\begin{figure}
\begin{center}
\includegraphics [width=15.5cm] {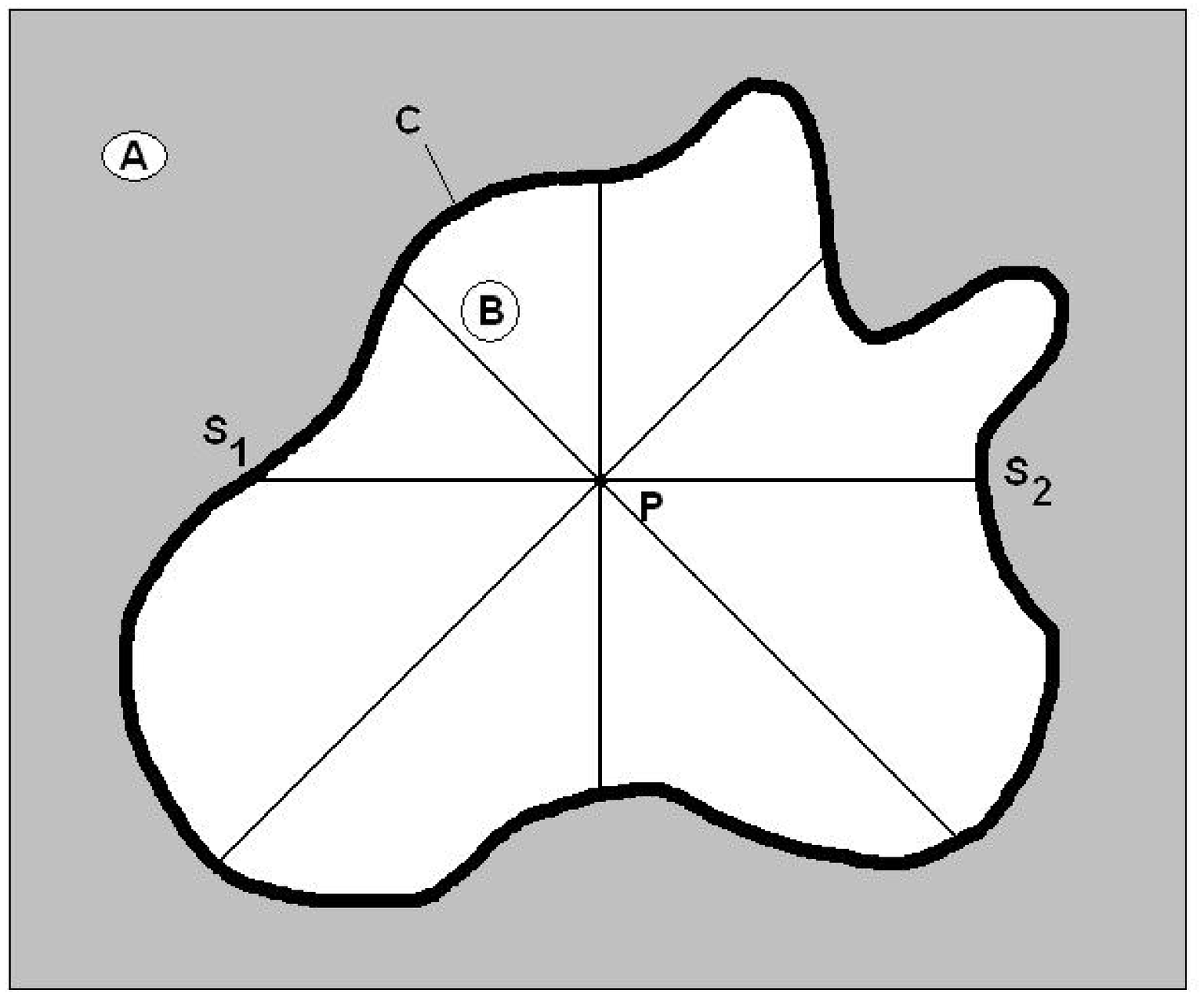}
\caption{\label{fig:Miniblock3}Evaluation of the ef\mbox{}fects of the BL/BSD at the point Q of F\mbox{}ig.~\ref{fig:Miniblock2} (not shown), 
whose projection onto the BL/BSD plane is the point P, on the basis of four directions (resulting, in this case, in four miniblocks). In this 
f\mbox{}igure, the point P lies within the area B.}
\end{center}
\end{figure}

To enable the evaluation of the ef\mbox{}fects of the BL/BSD at the point Q, a set of $N$ directions on the BL/BSD plane, intersecting 
at the point P, are chosen; an example with $N=4$ is shown in F\mbox{}ig.~\ref{fig:Miniblock3}. The ef\mbox{}fects of the BL/BSD at the point 
Q will be evaluated from the elementary contributions of the line segments corresponding to the intersection of $N$ straight lines 
with the contour C. These line segments, which are contained within the area B and are bound by the contour C, will be called 
\emph{miniblocks}. The ef\mbox{}fect of the BL/BSD at the point Q will be evaluated by averaging the elementary contributions of the 
miniblocks created around the point P. In F\mbox{}ig.~\ref{fig:Miniblock3}, one of these miniblocks is represented by the line segment 
$S_1 S_2$. Obviously, the number of miniblocks in each direction depends on the number of intersections of the straight line 
(drawn through the point P, parallel to the chosen direction) and the contour C.

\subsubsection{\label{sec:Without}The point P lies outside the area B}

\begin{figure}
\begin{center}
\includegraphics [width=15.5cm] {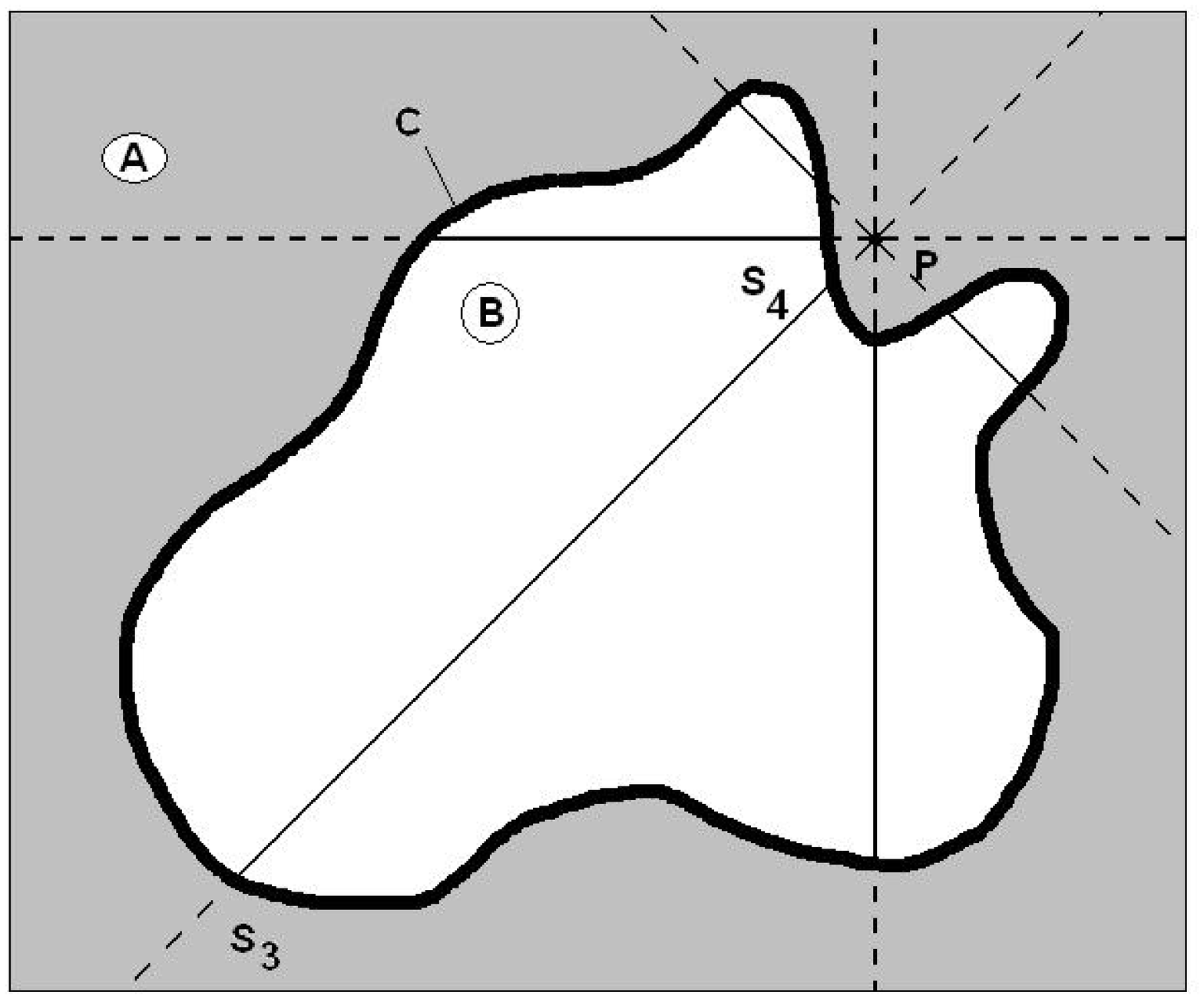}
\caption{\label{fig:Miniblock4}Evaluation of the ef\mbox{}fects of the BL/BSD at the point Q of F\mbox{}ig.~\ref{fig:Miniblock2} (not shown), 
whose projection onto the BL/BSD plane is the point P, on the basis of four directions (resulting, in this case, in f\mbox{}ive miniblocks). 
In this f\mbox{}igure, the point P lies outside the area B.}
\end{center}
\end{figure}

As in the previous section, a set of $N$ directions, intersecting at the point P, are chosen; an example with $N=4$ is shown 
in F\mbox{}ig.~\ref{fig:Miniblock4}. Again, the ef\mbox{}fects of the BL/BSD at the point Q will be evaluated on the basis of the elementary 
contributions in these $N$ directions. In F\mbox{}ig.~\ref{fig:Miniblock4}, one of the miniblocks is denoted by $S_3 S_4$. 

\subsubsection{\label{sec:Remarks}General remarks}

Evidently, the number of directions $N$ is arbitrary. For f\mbox{}ixed $N$, the accuracy of the evaluation depends on the details of the 
contour C and on the proximity of the point P to it. The reliability of the estimation is expected to increase with $N$. (The 
evaluation is exact for $N \to \infty$.)

There is one dif\mbox{}ference between the cases described in Sections \ref{sec:Within} and \ref{sec:Without}. In case that the point P 
lies within the area B, there will always be at least one miniblock per direction; if the point P lies outside the area B, there 
might be no intersections with the contour C in some directions (in which case, the corresponding elementary contributions vanish).

Let us assume that the physical characteristics of the beam (lateral spread, angular divergence, energy, etc.) and the entire 
geometry (also involving the BL/BSD) are f\mbox{}ixed. Apart from the original point Q, the elementary contributions will involve 
(the coordinates of) three additional points: the two end points of the particular miniblock and the point P. Due to the fact 
that the $N$ directions are created around the point P, the two end points of every miniblock and the point P will always lie on 
one straight line. If the point P lies within the area B, it may lie within or outside a miniblock. On the other hand, if the 
point P lies outside the area B, it will always lie outside any corresponding miniblock. In any case, the elementary contribution 
of a miniblock will generally be a function of the distance of its two end points to the point P. Additionally, the elementary 
contribution of a miniblock to the point Q will also involve the distance of the point Q to the BL/BSD plane, denoted as $z$ in 
F\mbox{}ig.~\ref{fig:Miniblock2}.

A few remarks are worth making.
\begin{itemize}
\item The angle between consecutive directions may be constant or variable. If the angle is not constant, weights (to the elementary 
contributions) have to be applied.
\item The values of $N$ in Sections \ref{sec:Within} and \ref{sec:Without} do not have to be the same. Furthermore, dif\mbox{}ferent 
values of $N$ might be used for the points which are projected onto the interior (or exterior) of the area B, e.g., depending on the 
proximity of the point P to the contour C.
\end{itemize}

At present, the method of this paper applies only to BL/BSDs with a constant aperture prof\mbox{}ile throughout their thickness; 
to derive the corrections in case of other shapes (e.g., for BL/BSDs with a tapered aperture prof\mbox{}ile), substantial modif\mbox{}ications 
are required. The approach is applicable to any type of BL/BSD, be it the primary collimator, the MLC, or the block; to derive the 
corrections, the only input pieces are the physical characteristics of the material (of which the BL/BSD is made) and, naturally, 
geometrical details. From now on, however, we will restrict ourselves to the ef\mbox{}fects induced by the block, which (given its 
proximity to the patient) are expected to be of greater interest and importance in clinical applications. This choice, meant to 
emphasise importance, should not be seen as a restriction of the method.

\subsection{\label{sec:Thickness}Block-thickness corrections}

A method for the evaluation of the thickness corrections was recently proposed by Slopsema and Kooy (2006); we will follow their 
terminology. Thus, `downstream (upstream) projection' will indicate the projection of the downstream (upstream) face of the block 
on the ($x$,$y$) plane at the specif\mbox{}ied $z$ position (depth). Similarly, the `extension' of the block will correspond to 
the physical block translated in space (from its actual position to the specif\mbox{}ied depth, parallel to the central beam axis).

\begin{figure}
\begin{center}
\includegraphics [width=15.5cm] {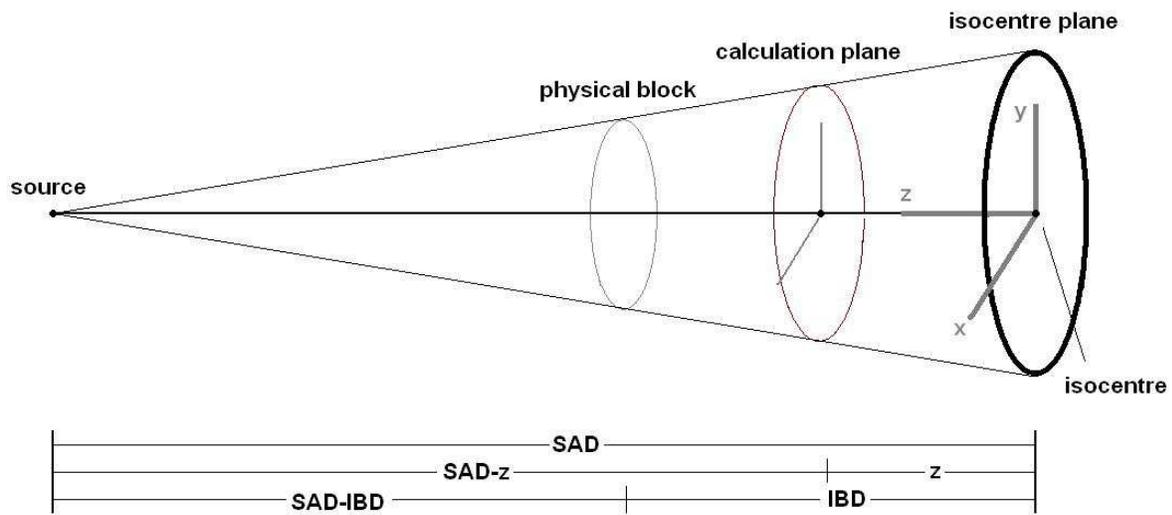}
\caption{\label{fig:Geometry1}The relation between the DICOM block, the extension, and the downstream projection on the ($x$,$y$) plane 
(calculation plane) at a specif\mbox{}ied depth $z$.}
\end{center}
\end{figure}

In medical applications, the formal beam-shaper object (e.g., the so-called DICOM block), which is created at planning time to represent 
the physical block, comprises the projection of the \emph{downstream face} (of the block) to a plane (perpendicular to the beam) at 
isocentre depth. The physical block (extension) is therefore obtained via a simple scaling involving the source-axis and block-isocentre 
(IBD) distances. Having retrieved the extension of the block, it is straightforward to obtain the downstream projection at any $z$; in 
order to obtain the upstream projections, one has to use, in addition to the aforementioned quantities, the block thickness ($d$). The 
relation between the DICOM block, the extension, and the downstream projection at a specif\mbox{}ied $z$ position is shown in 
F\mbox{}ig.~\ref{fig:Geometry1}. The upstream projection is obtained by projecting the extension from depth $z={\rm IBD}+d$. The thin-block 
approximation, which is currently used in clinical applications when evaluating the dose, corresponds to $d=0$ mm (the downstream and 
upstream projections coincide at all $z$ values).

Assuming that the coordinates of the projected (downstream- or upstream-face, as the case might be) ends of a miniblock are denoted as 
$x_1$ and $x_2$, the contribution of the miniblock to the f\mbox{}luence at a point $x_p$ (lying on the straight line def\mbox{}ined by 
$x_1$ and $x_2$) on the calculation plane is given by the formula
\begin{equation} \label{eq:1DContributions}
F(x_p)=\frac{1}{2} \mid {\rm erf}(\frac{x_p-x_1}{\sigma_1 \sqrt{2}}) - {\rm erf}(\frac{x_p-x_2}{\sigma_2 \sqrt{2}}) \mid \, ,
\end{equation}
where ${\rm erf}(x)$ denotes the error function and $\sigma_{1,2}$ stand for the rms (lateral) spreads of the beam at the specif\mbox{}ied 
depth. The quantities $\sigma_{1,2}$ are obtained via source mirroring according to the method of Slopsema and Kooy (2006); as points 
on the downstream or upstream faces of the block are used, the resulting $\sigma$ values in Eq.~(\ref{eq:1DContributions}) are equal only 
when one block face is involved in the mirroring process.

\begin{figure}
\begin{center}
\includegraphics [width=15.5cm] {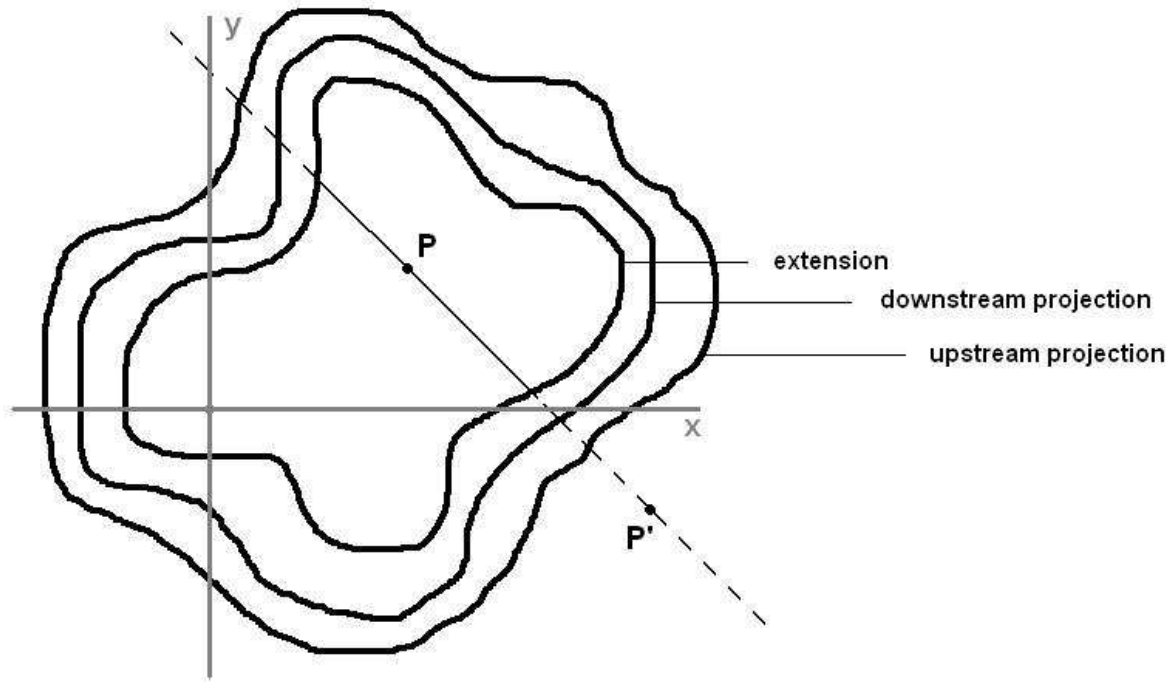}
\caption{\label{fig:Geometry2}The extension and the projections of the downstream and upstream faces of the block on a calculation plane. 
The point P lies within the extension, the point P$^\prime$ without.}
\end{center}
\end{figure}

One simple example of the projections on the calculation plane is shown in F\mbox{}ig.~\ref{fig:Geometry2}; the block extension is contained 
within the downstream projection, which (in turn) is contained within the upstream projection. In reality, depending on the complexity 
of the shape of the block aperture and on the relative position of the central beam axis, these three contours might intersect one another. 
The central beam axis intersects the calculation plane at the origin of the ($x$,$y$) coordinate system. One miniblock is shown (the line 
segment contained within the block extension), along with two points, one within the miniblock (P), another outside the miniblock 
(P$^\prime$). The f\mbox{}luence contributions to both points may be evaluated by using Eq.~(\ref{eq:1DContributions}) with the appropriate 
$x_{1,2}$ and $\sigma_{1,2}$ values.

\begin{figure}
\begin{center}
\includegraphics [width=15.5cm] {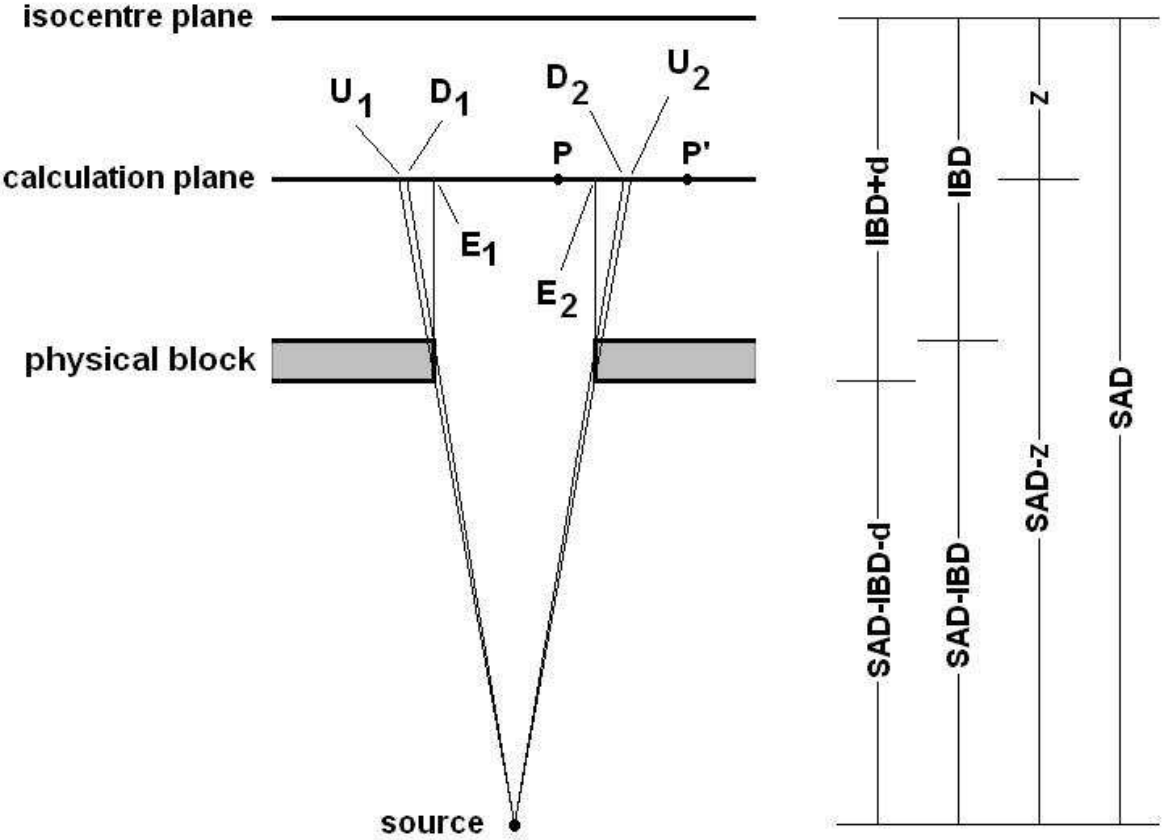}
\caption{\label{fig:Geometry3}The essentials for the evaluation of the contribution of a miniblock to the f\mbox{}luence at specif\mbox{}ied 
points (e.g., at P and P$^\prime$) on the calculation plane.}
\end{center}
\end{figure}

The essentials for the evaluation of the contribution of a miniblock to the f\mbox{}luence at a specif\mbox{}ied point on the calculation plane 
are to be found in F\mbox{}ig.~\ref{fig:Geometry3}. Although, in the general case, the miniblock does not contain the intersection of the $x$ 
and $y$ axes of F\mbox{}ig.~\ref{fig:Geometry2}, all lengths (which are important to our purpose) scale by the same factor, thus enabling the 
simplif\mbox{}ied picture of F\mbox{}ig.~\ref{fig:Geometry3}.

According to Slopsema and Kooy (2006), the source is mirrored onto the calculation plane by using points either on the downstream or on the 
upstream face of the block. The values of the mirrored-source size, corresponding to these two options, are given by
\begin{equation} \label{eq:SigmaD}
\sigma_d=\sigma_s \frac{{\rm IBD}-z}{{\rm SAD}-{\rm IBD}} \,
\end{equation}
and
\begin{equation} \label{eq:SigmaU}
\sigma_u=\sigma_s \frac{{\rm IBD}-z+d}{{\rm SAD}-{\rm IBD}-d} \, .
\end{equation}

The coordinates of the projected points $U_1$, $D_1$, $D_2$, and $U_2$ are obtained on the basis of F\mbox{}ig.~\ref{fig:Geometry3} via 
simple operations. The last item needed to derive the contributions to the f\mbox{}luence may be found on page 5444 of the paper of Slopsema 
and Kooy (2006): `Protons whose tracks project inside the aperture extension onto the plane of interest see the upstream face as the 
limiting boundary. Only for protons whose tracks end up outside the aperture on the plane of interest is the downstream face the limiting 
aperture boundary.' In fact, this statement applies to the case of a half-block (one-sided block). The modif\mbox{}ication, however, in 
case of a miniblock is straightforward.
\begin{itemize}
\item Points within the extension of the aperture see the upstream face of the miniblock as the limiting boundary.
\item Points outside the extension of the miniblock see one upstream \emph{and one downstream} edge as limiting boundaries.
\end{itemize}
Obviously, in order to evaluate the f\mbox{}luence at the point P (F\mbox{}ig.~\ref{fig:Geometry3}), one has to use the coordinates 
of the projected points $U_1$ and $U_2$, along with $\sigma_u$ in Eq.~(\ref{eq:1DContributions}). On the other hand, to evaluate the 
f\mbox{}luence at P$^\prime$, one has to use $U_1$ along with $\sigma_u$ (contribution of the `left' part of the miniblock) 
and $D_2$ along with $\sigma_d$ (contribution of the `right' part of the miniblock). This simplif\mbox{}ied picture, featuring what a 
point `perceives' as limiting boundaries, suf\mbox{}f\mbox{}ices in obtaining the appropriate f\mbox{}luence contributions.

\begin{figure}
\begin{center}
\includegraphics [width=15.5cm] {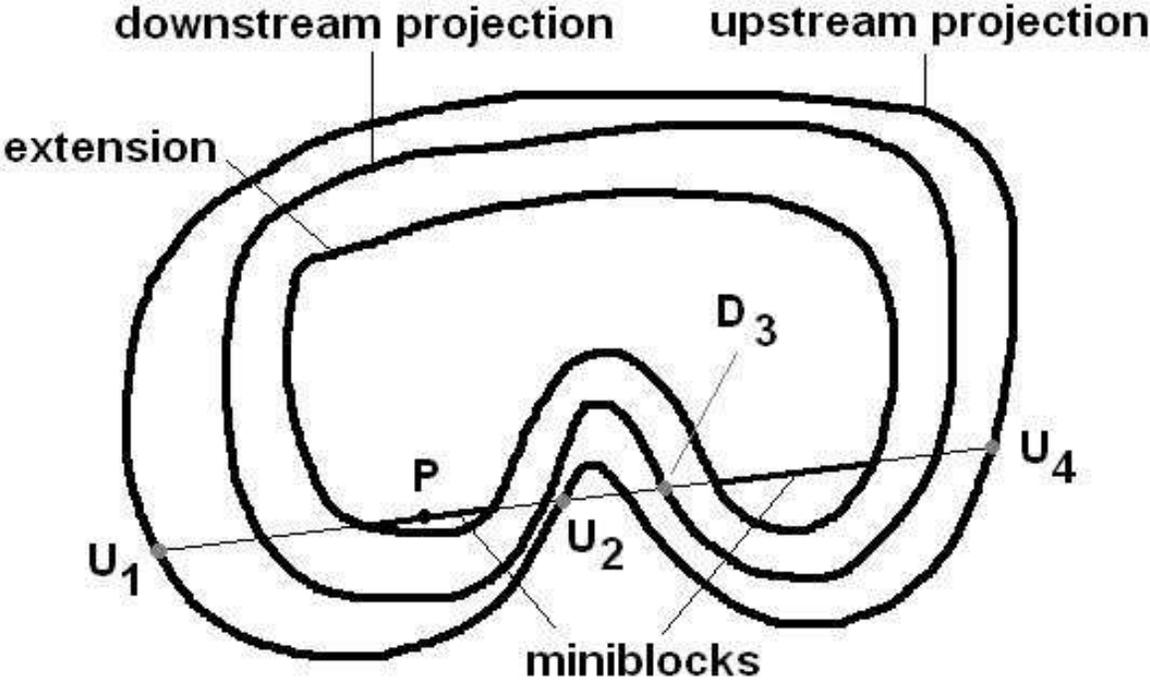}
\caption{\label{fig:Geometry4}Due to one-dimensional nature of the miniblocks, a point lying within the extension (of a miniblock) may also 
receive contributions which are characteristic to points lying in the exterior of the extension.}
\end{center}
\end{figure}

F\mbox{}inally, the number of contributions which a point will receive depends on the geometry and on the shape of the block. One example 
of a point receiving two contributions in a given direction is shown in F\mbox{}ig.~\ref{fig:Geometry4}; the point P lies within the 
extension of the miniblock on the left and outside the extension of the miniblock on the right. Additionally, one might have to deal with 
a block with more than one apertures (hence contours). To determine in the present work the appropriate thickness corrections for arbitrary 
block-aperture shape and beamline geometry, dedicated software was developed.

\subsection{\label{sec:Scattering}Block-scattering corrections}

Before entering the details of the derivation of the scattering corrections, it is worth providing a concise outline of our approach. 
F\mbox{}irst of all, our aim is to obtain a \emph{fast and reliable} solution to the scattering problem; exact analytical solutions 
are welcome as long as they fulf\mbox{}ill this requirement. Second, the solution has to be general enough for direct application to 
all proton-treatment machines.

To expedite the derivation and application of the corrections at planning time, we will introduce a two-step approach.
\begin{itemize}
\item All the parameters which are independent of the specif\mbox{}ic details of plans will be evaluated during the beam-conf\mbox{}iguration 
phase; given a f\mbox{}ixed (hardware) setup, every treatment machine is conf\mbox{}igured only once.
\item The corrections for each particular plan will be derived (at planning time) from the existing results (i.e., those obtained at 
beam-conf\mbox{}iguration phase) via simple interpolations.
\end{itemize}

Given that the physics of multiple scattering is known, it is possible to obtain the exact solution for the relevant beam properties 
(lateral spread, angular divergence) from the beamline characteristics of each treatment machine. However, it is unrealistic to introduce 
a dedicated process for each supported machine, especially in order to derive corrections to the delivered dose. Additionally, if a 
dedicated (per-case) approach is implemented in a software product which is intended to support a variety of machine manufacturers, one 
has to be prepared to allot the necessary resources whenever a new product appears. To avoid these problems and to retain the generality 
of the approach, one has no other choice but to introduce a simple, adjustable model to account for the beam optics. The model of the 
present paper has only one parameter, which will be f\mbox{}ixed from half-block f\mbox{}luence measurements.

The parameters which achieve the description of the various distributions of the scattered protons are determined via MC runs. These runs 
take account of the variability in the block material, block thickness, incident energy (energy at nozzle entrance), and nozzle-equivalent 
thickness (NeT) for all the options (combinations of the hardware components of the beamline, leading to ranges of available energies and 
of NeTs, as well as imposing restrictions on the f\mbox{}ield size) for which a machine is conf\mbox{}igured. To enable the easy use of 
the results, the output is put in the form of expansion parameters in two geometrical quantities which are involved in the description of 
the scattering ef\mbox{}fects.

The scattering corrections for all the blocks in a plan are determined (at planning time) from the aforementioned results via simple 
interpolations. The application of the corrections involves the concept of miniblocks, as they have been introduced in Section 
\ref{sec:Miniblocks} of this paper.

\subsubsection{\label{sec:BeamModelling}Modelling of the beam}

One model which is frequently used in beam optics features the bivariate Gaussian distribution in the lateral direction $y$ (distance to 
the central beam axis) and the (small) angle $\theta$ (with respect to that axis). (Rotational symmetry is assumed here.)
\begin{equation} \label{eq:Bivariate}
f(y,\theta) = \frac{1}{\pi \sqrt{D}} \, {\rm exp} \Big( - \frac{A \theta^2 - 2 B y \theta + C y^2}{D} \Big) \, ,
\end{equation}
with
\begin{equation} \label{eq:D}
D = A C - B^2 \, .
\end{equation}
The parameters $A$ and $C$ represent twice the variance in $y$ and $\theta$, respectively. The ($y$,$\theta$) correlation is def\mbox{}ined as
\begin{equation} \label{eq:Rho}
\rho = \frac{B}{\sqrt{A C}} \, .
\end{equation}
The quantity $\rho$ (which is bound between $-1$ and $1$) is a measure of the focusing in the beam. Positive $\rho$ values indicate a 
defocusing system, negative a focusing one; this becomes obvious after one puts Eq.~(\ref{eq:Bivariate}) in the form
\begin{equation} \label{eq:BivariateNew}
f(y,\theta) = \frac{1}{\sigma_\theta \sqrt{2 \pi}} \, {\rm exp} \Big( - \frac{\theta^2}{2 \sigma^2_\theta} \Big) \frac {1}{\sigma_s \sqrt{2 \pi (1-\rho^2)}} 
\, {\rm exp} \Big( - \frac{\Big( y - \rho \frac{\sigma_s}{\sigma_\theta} \theta \Big)^2}{2 \sigma^2_s (1-\rho^2)} \Big) \, .
\end{equation}

We now touch on the variation of $A$, $B$, and $C$ along the beam-propagation direction. Assuming that the quantities $A_0$, $B_0$, 
and $C_0$ denote the corresponding values at isocentre depth and that the beam propagates in air (without scattering), $A$, $B$, and 
$C$ at distance $z$ from the isocentre (see F\mbox{}ig.~\ref{fig:Geometry1}) are given by the expressions
\begin{equation} \label{eq:ABCz}
A(z) = A_0 - 2 B_0 z + C_0 z^2 \,\, , B(z) = B_0 - C_0 z \,\, , {\rm and} \,\, C(z) = C_0 \, .
\end{equation}
With these transformations, the joint probability distribution of Eq.~(\ref{eq:Bivariate}) is invariant under translations in $z$. In 
case that the beam traverses some material, Eqs.~(\ref{eq:ABCz}) have to be modif\mbox{}ied accordingly, to take account of the beam 
broadening due to multiple scattering.

\subsubsection{\label{sec:Parameterisation}Simplif\mbox{}ied parameterisation of the beamline}

The accurate modelling of the beam may be obtained on the basis of formulae such as those given in the previous section. In Section 
\ref{sec:Scattering}, however, we reasoned that a simplif\mbox{}ied parameterisation of the beamline is desirable; one additional 
argument may be put forth.

Currently, as far as proton therapy is concerned, four dose-delivery techniques are in use: single-scattering, double-scattering, 
uniform-scanning (formerly known as wobbling), and modulated-scanning (formerly simply known as scanning). In the modulated-scanning 
technique, magnets def\mbox{}lect a narrow beam onto a sequence of pre-established points (spots) on the patient (for pre-determined optimal 
times), thus `scanning' the (cross section of the) region of interest. Uniform scanning involves the spread-out of the beam using fast 
magnetic switching. The broadening of the beam in the single-scattering technique is achieved by one scatterer, made of a high-$Z$ 
material and placed close to the entrance of the nozzle. Currently, the most `popular' technique involves a double-scattering system.

In a double-scattering system, a second scatterer is placed downstream of the f\mbox{}irst scatterer in order to achieve ef\mbox{}f\mbox{}icient 
broadening of the beam; studies of the ef\mbox{}fects of the second scatterer may be found in the literature, e.g., see Takada (2002) and more 
recent reports by the same author and Gottschalk. The second scatterer is usually made of two materials: a high-$Z$ (such as lead) material 
at the centre (i.e., close to the central beam axis), surrounded by a low-$Z$ (such as aluminium, lexan, etc.) material (which is 
frequently, but not necessarily, shaped as a concentric ring). The arrangement produces more scattering at the centre than the 
periphery, leading (after sophisticated f\mbox{}ine-tuning) to the creation of a broad f\mbox{}lat f\mbox{}ield at isocentre.

To simulate the ef\mbox{}fect of the second scatterer in the present work, ($y$,$\theta$) events are generated at $z={\rm SAD}$ as follows. 
The variable $y$ is sampled from the Gaussian distribution with mean $0$ and variance $\sigma_s^2$ (the source-size calibration must precede 
this step); $\sigma_s$ depends on the incident energy and NeT. To account for the lateral limits of the beam, $y$ is restricted within the 
interval $[-R_L,R_L]$, where the characteristic length $R_L$ is taken herein to be the radius of the second scatterer. The variable $\theta$ 
is f\mbox{}irst sampled from the Gaussian distribution with mean $0$ and a $y$-dependent variance $\sigma_\theta^2$ according to the formula
\begin{equation} \label{eq:SigmaTheta}
\sigma_\theta (t) = \sigma_\theta(1) \Big( (1-\lambda) \mid t \mid + \lambda \Big) \, ,
\end{equation}
where $t$ denotes the lateral position as a fraction of $R_L$; $\mid t \mid = \mid y/R_L \mid \leq 1$. To account for the $y$-dependent 
bias in $\theta$, we then use the transformation $\theta \rightarrow \theta + {\rm tan}^{-1} (y/L) \simeq \theta + y/L$, where $L$ stands 
for the distance between the f\mbox{}irst scatterer and the source. Obviously, $\lambda$ denotes the ratio of two $\sigma_\theta$ values, i.e., 
the value at the centre of the second scatterer over the one at the rim; $\lambda$ is the only free parameter of the model. The value of 
$\sigma_\theta(1)$ is obtained from the incident energy and NeT; the angular divergence of the beam at nozzle entrance is (currently) 
assumed to be $0$.

It is not dif\mbox{}f\mbox{}icult to prove that, given the aforementioned rules for generating the ($y$,$\theta$) events, the open-f\mbox{}ield 
f\mbox{}luence at a lateral position $y$ at depth $z<{\rm SAD}$ is given by
\begin{equation} \label{eq:FluenceOpen}
f(y) = \frac{R_L}{2 \pi \sigma_s ({\rm SAD}-z)} \, \int_{-1}^{t_0} \frac{dt}{\sigma_\theta (t)} \, {\rm exp} \Big( - \frac{t^2R_L^2}{2 \sigma_s^2} \Big) 
{\rm exp} \Big( - \frac{( \frac{y-t R_L}{{\rm SAD}-z} - \frac{tR_L}{L})^2}{2 \sigma_\theta^2 (t)} \Big) \, ,
\end{equation}
where $t_0=1$. It has to be emphasised that this def\mbox{}inition of the f\mbox{}luence does not involve an overall $\frac{1}{r^2}$-factor 
($r$ being the distance between the calculation plane and the source); therefore, this formula is compatible with the format in 
which the lateral f\mbox{}luence measurements, used during the beam-conf\mbox{}iguration phase, appear.

It can be shown that the only modif\mbox{}ication in case that a half-block is inserted into the beamline (e.g., as shown in 
F\mbox{}ig.~\ref{fig:Model1}) involves the upper limit $t_0$ of integration in Eq.~(\ref{eq:FluenceOpen}); instead of $t_0=1$, 
one must now use
\begin{equation} \label{eq:t0}
t_0 = {\rm min} \left\{ \frac{b ({\rm SAD}-z) - y ({\rm SAD}-{\rm IBD})}{R_L ({\rm IBD}-z)} , 1 \right\}
\end{equation}
if $t_0 > -1$, or otherwise $t_0=-1$.

Although the method of the present paper was originally developed for the double-scattering technique, it is also applicable in single 
scattering and uniform scanning; in both cases, one simply has to f\mbox{}ix the parameter $\lambda$ to $1$. As the method is applicable 
only in case of broad f\mbox{}ields, it has no bearing on modulated scanning.

\subsubsection{\label{sec:Moliere}Multiple scattering through small angles}

The elements needed for the description of the passage of particles through matter may be found (in a concise form) in Yao \etal 
(2006), starting on page 258; most of the def\mbox{}lection of a charged particle traversing a medium is due to Coulomb scattering of\mbox{}f 
nuclei. Despite its incompleteness (e.g., see the discussion in the GEANT4 physics-reference manual, section on `Multiple Scattering', 
starting on page 71), the multiple-scattering model of Moli{\`e}re is used here. The large-angle scattering is not taken into 
account; the angular distribution of the traversing beam is assumed Gaussian.

Highland's logarithmic term, appearing in the expression of $\theta_0$, will be approximated by a constant factor involving 
the block thickness; a similar strategy was followed in Gottschalk \etal (1993). The Lynch-and-Dahl (1991) values will 
be used in the expression for $\theta_0$:
\begin{equation} \label{eq:Theta0}
\theta_0(q) = \frac{13.6MeV}{\beta c p} \sqrt{\frac{q}{X_0}} \Big( 1 + 0.038 \, {\rm ln} ( \frac{d}{X_0} ) \Big) \, ,
\end{equation}
where $q$ denotes the depth along the original trajectory, $\beta c$ the velocity and $p$ the momentum of the proton, and $X_0$ 
the radiation length in the material of the block (for a convenient parameterisation of $X_0$, see Yao \etal (2006), page 263).

Equation (\ref{eq:Theta0}) applies to `thin' targets. For `thick' targets, the dependence of the proton momentum on the 
depth $q$ must be taken into account. Omitting the logarithmic term on the right-hand side, one may put 
Eq.~(\ref{eq:Theta0}) in the form
\begin{equation} \label{eq:Theta0Simple}
\theta_0(q) = 2 f(q) \sqrt{q} \, ,
\end{equation}
where
\begin{equation} \label{eq:Fq}
f(q) = \frac{13.6MeV}{2 \sqrt{X_0}} \Big( \frac{1}{q} \int_0^q \frac{dt}{(\beta(t) c p(t))^2} \Big)^{1/2} \, .
\end{equation}
On the practical side, Eq.~(\ref{eq:Theta0Simple}) with a constant (or at least `not too complicated') $f(q)$ factor would be attractive 
as one would then be able to obtain fast analytical solutions for the propagation of a simulated track inside the material of the block. 
Therefore, it is worth determining the range of thickness values within which the constancy of the $f(q)$ factor remains a reasonable 
assumption. The direct comparison with the data of Gottschalk \etal (1993) revealed that the `thick-target' corrections are unfortunately 
indispensable at depths exceeding about $50 \%$ of the range of $158.6$-MeV protons, incident on a variety of materials.

To abide by the original goal of obtaining a fast solution, we had to follow an alternative approach (to that of using Eq.~(\ref{eq:Fq})), 
by parameterising the $q$-dependence of the $f(q)$ factor in a simple manner; at present, the best choice seems to be to make use of the 
empirical formula
\begin{equation} \label{eq:Theta0New}
\theta_0(q) = \frac{2 f \sqrt{q}}{1-\frac{q^2}{1.75 R^2}} \, ,
\end{equation}
where $R$ denotes the (energy-dependent) range of the incident proton in the material of the block and
\begin{equation} \label{eq:F}
f = \frac{13.6MeV}{\beta c p} \frac{1}{2 \sqrt{X_0}} \Big( 1 + 0.038 \, {\rm ln} ( \frac{d}{X_0} ) \Big) \, ;
\end{equation}
$f$ is now a constant, depending only on the initial value of $\beta$. The validation of Eqs.~(\ref{eq:Theta0New}) and (\ref{eq:F}) was 
made on the basis of a comparison with the experimental data of Gottschalk \etal (1993), namely the measured $\theta_M$ values of that 
paper. Good agreement with the data was obtained for four materials which are of interest in the context of the present study (carbon, 
lexan, aluminum, and lead). Only at one entry (one of the largest depth values in lead, i.e., the measurement at $q/R=0.97548$), was a 
signif\mbox{}icant dif\mbox{}ference (of slightly less than $20 \%$) found; the origin of that dif\mbox{}ference was not sought.

\subsubsection{\label{sec:MCruns}Details on the generation of the MC events}

\begin{figure}
\begin{center}
\includegraphics [width=15.5cm] {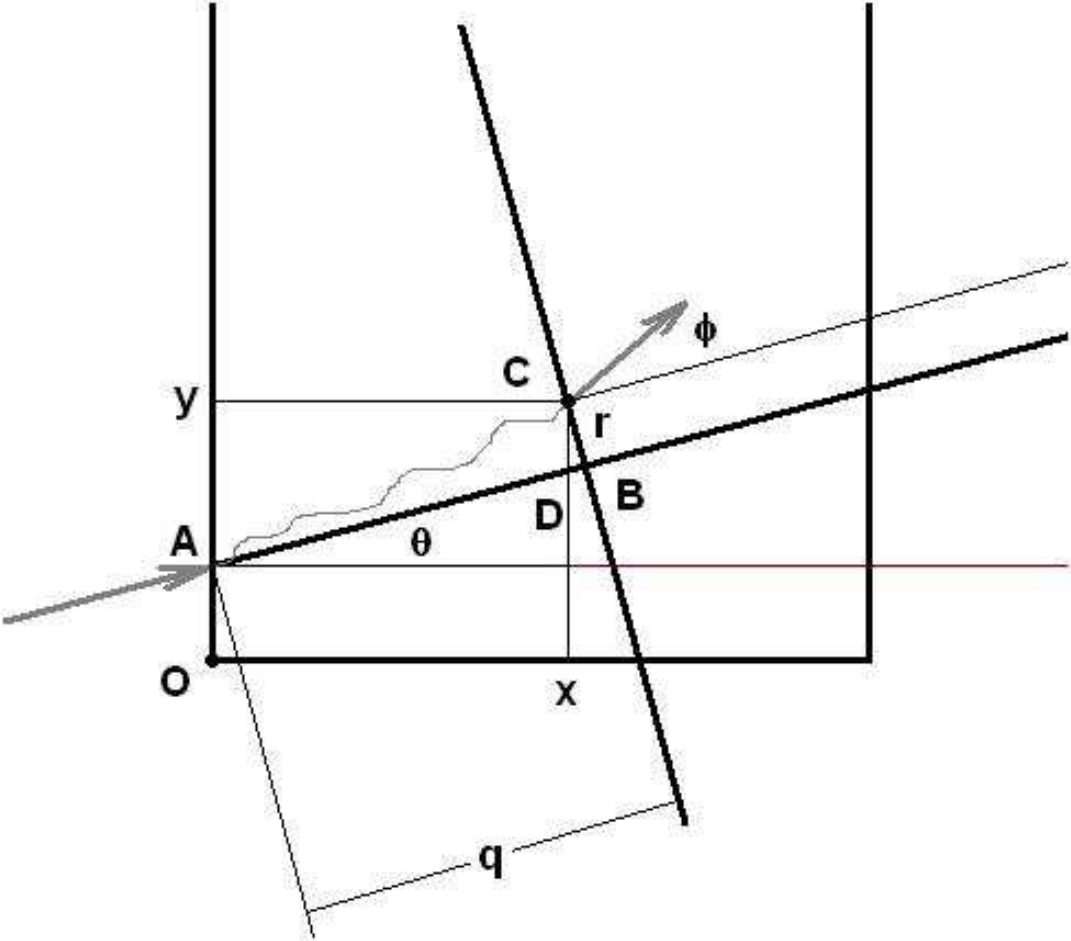}
\caption{\label{fig:Model2}The description of the kinematics inside the block. The auxiliary coordinate system introduced in this 
f\mbox{}igure should not be confused with the formal coordinate system of F\mbox{}ig.~\ref{fig:Geometry1}.}
\end{center}
\end{figure}

F\mbox{}igure \ref{fig:Model2} shows an example of a trajectory of one proton inside a block. The incident proton, an OT in this 
f\mbox{}igure, hits the upstream face of the block at angle $\theta$ with respect to the beam axis. Two new \emph{in-plane} variables 
are introduced to describe the kinematics at depth $q$ (along the direction of the original trajectory): the def\mbox{}lection $r$ 
(of\mbox{}f the original-trajectory course) and the angle $\phi$ (with respect to the direction of the original trajectory). Although 
the proton moves in an irregular path inside the block, the `history' of the actual motion will be replaced by a smooth movement leading 
to the same value of $r$ at $q$. This `smooth-def\mbox{}lection' approximation will enable the association of the energy loss in the material 
of the block with the doublet of ($q$,$r$) values. Since the path length, which is calculated in this approximation, is an underestimate 
of the actual path length, a constant~\footnote{In reality, the true-path correction is expected to be energy-dependent.} conversion factor 
(true-path correction) of $0.9$ has been used; there is some arbitrariness concerning this choice, yet it appears to be reasonable. It 
needs to be stressed that, in F\mbox{}ig.~\ref{fig:Model2}, $x$ denotes the direction of the beam propagation; the auxiliary coordinate 
system introduced in this f\mbox{}igure should not be confused with the coordinate system of F\mbox{}ig.~\ref{fig:Geometry1}, which is 
the formal one in medical applications.

In the generation of the MC events, the suggestion of Yao \etal (2006), page 262, for the quantities $r$ and $\phi$ has been followed. 
Two independent Gaussian random variables ($z_1$,$z_2$) with mean $0$ and variance $1$ are f\mbox{}irst created in each event (track hitting 
the block). The quantities $r$ and $\phi$ are expressed in terms of ($z_1$,$z_2$) as follows:
\begin{equation} \label{eq:Rq}
r(q) = \Big( \frac{z_1}{\sqrt{3}} + z_2 \Big) \frac{q \theta_0(q)}{2}
\end{equation}
and 
\begin{equation} \label{eq:Phiq}
\phi(q) =z_2 \theta_0(q) \, ,
\end{equation}
where $\theta_0(q)$ is taken from Eq.~(\ref{eq:Theta0New}).

The values of the doublet ($z_1$,$z_2$) f\mbox{}ix the dependence of $r$ and $\phi$ on $q$ in each generated event. It may then be determined 
(either analytically or numerically) whether the particular track leaves the block. F\mbox{}inally, for those tracks leaving the block, 
a simple rotation yields the coordinates of the exit point in the ($x$,$y$) coordinate system of F\mbox{}ig.~\ref{fig:Model2}. The energy of 
the leaving proton is determined from its residual range (original range minus the actual path length inside the material of the block).

\subsubsection{\label{sec:LatDist}The lateral f\mbox{}luence distributions of the scattered protons}

\begin{figure}
\begin{center}
\includegraphics [width=15.5cm] {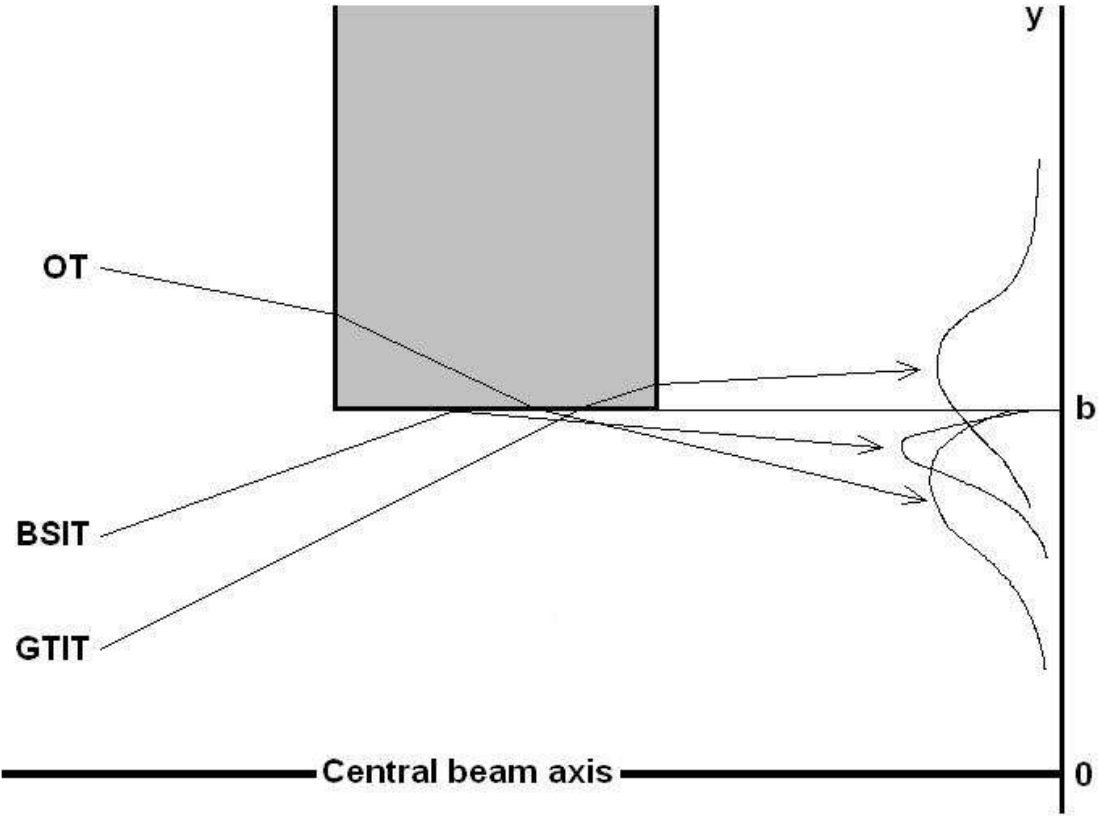}
\caption{\label{fig:Model1}The outer tracks (OTs) and the bore-scattered inner tracks (BSITs) emerge from the bore. The going-though inner 
tracks (GTITs) emerge from the downstream face of the block.}
\end{center}
\end{figure}

The def\mbox{}initions of the three types of scattered protons have been given in Section \ref{sec:Introduction}; in F\mbox{}ig.~\ref{fig:Model1}, 
one obtains a rough schematic view of the corresponding contributions to the f\mbox{}luence. In this section, we will introduce convenient 
forms to parameterise the lateral f\mbox{}luence distributions of the scattered protons.

As far as these distributions are concerned, the uninteresting of\mbox{}fset $b$ (lateral displacement of the block) in 
F\mbox{}ig.~\ref{fig:Model1} will be omitted. Therefore, for the needs of this section, $y=0$ mm at the extension of the 
block in F\mbox{}ig.~\ref{fig:Model1}, that is, not at the position where the central beam axis intersects the calculation plane. 
To obtain the lateral f\mbox{}luence distributions, literally corresponding to F\mbox{}ig.~\ref{fig:Model1}, one has to 
take the of\mbox{}fset $b$ into account. In the formulae below, the placement of the half-block is assumed as shown in 
F\mbox{}ig.~\ref{fig:Model1} (i.e., extending to $+ \infty$).

The empirical formula for the description of the lateral f\mbox{}luence distribution of the OTs reads as
\begin{equation} \label{eq:OT}
f(y) = \alpha \, {\rm exp} (-y / \beta) \, (y + \gamma)^2 \, ,
\end{equation}
where $y \leq 0$. Four conditions for the parameters $\alpha$, $\beta$, and $\gamma$ must be fulf\mbox{}illed: $\alpha>0$, $\beta < 0$, 
$\gamma \leq 0$, and $2 \beta - \gamma \leq 0$.

The same empirical formula is used in the parameterisation of the lateral f\mbox{}luence distribution of the BSITs; the four aforementioned 
conditions also apply. (The resulting optimal parameter values are, of course, dif\mbox{}ferent.)

The optimal description of the (broader distribution of the) GTITs is achieved on the basis of a Lorentzian
\begin{equation} \label{eq:GTIT}
f(y) = \frac{\alpha \gamma^2}{(y-\beta)^2+\gamma^2} \, ,
\end{equation}
multiplied by the asymmetry factor
\[
\frac{2}{1+{\rm exp}(-2 \frac{y-\beta}{\gamma})} \, ,
\]
to account for the observed skewness of the lateral f\mbox{}luence distribution (toward positive $y$ values). Three conditions must be 
fulf\mbox{}illed: $\alpha>0$, $\beta \geq 0$, and $\gamma>0$.

The lateral f\mbox{}luence distribution of the pristine tracks is f\mbox{}itted by using the standard formula
\begin{equation} \label{eq:Pristine}
f(y) = \alpha \Big( 1 - \Phi (\frac{y - \beta}{\gamma}) \Big) \, , 
\end{equation}
where $\alpha>0$ and $\gamma>0$; $\Phi(x)$ is the cumulative Gaussian distribution function. Only the factor $\alpha$ will be retained 
from the f\mbox{}its to the pristine-beam data, to be used in the \emph{normalisation} of the f\mbox{}luence corresponding to each of 
the three types of scattered protons. Expressing the contributions of the scattered protons as fractions of the pristine-beam 
f\mbox{}luence enables the ef\mbox{}f\mbox{}icient application of the corrections at planning time.

From all the above, it is evident that the description of the contribution to the f\mbox{}luence (at f\mbox{}ixed $z$) of any of the 
three types of the scattered protons is achieved on the basis of three parameters; hence, there are nine parameters in total. At the 
end of each cycle (comprising a set of MC runs for a number of $b$ values, for a given energy-NeT combination), 
each of these nine parameters $p_i$ is expanded in terms of $b$ and $z$, using the quadratic model
\begin{equation} \label{eq:SR} 
p_i = a_{i1} + a_{i2} b + a_{i3} z + a_{i4} b z + a_{i5} b^2 + a_{i6} z^2 \, .
\end{equation}
The f\mbox{}inal results are the $56$ coef\mbox{}f\mbox{}icients $a_{ij}$, $i \in \{ 1, \dots , 9 \}$, $j \in \{ 1, \dots , 6 \}$, 
obtained at several energy-NeT combinations for the option of the machine which is under calibration. At planning time, the values 
of the parameters $p_i$ are reconstructed from the existing results (via simple 2D interpolations in the incident energy and NeT), 
the values of $b$ (corresponding to the particular miniblock which is processed), and $z$ (corresponding to the calculation plane 
which is processed).

\subsubsection{\label{sec:Outline}A summary of the approach}

Let us assume that one option of a selected machine has been chosen for calibration (at beam-conf\mbox{}iguration phase). All half-block data 
(which is contained in that option) is used in the determination of the parameter $\lambda$ on the basis of an optimisation scheme (featuring 
the C++ implementation of the standard optimisation package MINUIT of the CERN library), along with Eq.~(\ref{eq:FluenceOpen}) with the $t_0$ 
value of Eq.~(\ref{eq:t0})~\footnote{For the determination of the parameter $\lambda$, one could also use the measurements of the 
open-f\mbox{}ield f\mbox{}luence, along with Eq.~(\ref{eq:FluenceOpen}) with $t_0=1$. However, what is habitually called `open-f\mbox{}ield 
measurements' in the f\mbox{}ield of radiation therapy corresponds to beams which have already been restricted in size by the primary collimator.}.

Representative incident-energy and NeT values are chosen from the ranges of values associated with the option which is under calibration. 
For each acceptable energy-NeT combination, a number of MC runs are performed, each corresponding to one $b$ value. In each of these MC 
runs, events (i.e., ($y$,$\theta$) pairs, each corresponding to one proton track) are generated according to the formalism developed in 
Section \ref{sec:Parameterisation}. The value of the parameter $\lambda$, obtained from the half-block data (for the option in question) 
at the previous step, comprises (i.e., apart from geometrical characteristics of the machine which is conf\mbox{}igured) the only input 
to these MC runs. The resulting tracks are followed until they either hit the block or pass through it. The tracking of the protons inside 
the material of the block is done according to Section \ref{sec:MCruns}; f\mbox{}inally, these tracks either vanish (being absorbed in the 
material of the block) or emerge from it (bore, downstream face) and deliver dose.

The tracks which emerge from the block are properly f\mbox{}lagged (OTs, BSITs, or GTITs) and their contributions to the f\mbox{}luence at 
a number of $z$ depths are stored (histogrammed). F\mbox{}its to these distributions, using the empirical formulae of Section \ref{sec:LatDist}, 
lead to the extraction of the parameters achieving the optimal description of the stored data. After the completion of all the runs for all 
the chosen values of $b$, the entire set of the parameter values is subjected to f\mbox{}its by using Eq.~(\ref{eq:SR}) for each parameter 
separately. F\mbox{}inally, the coef\mbox{}f\mbox{}icients $a_{ij}$, $i \in \{ 1, \dots , 9 \}$, $j \in \{ 1, \dots , 6 \}$, appearing 
in Eq.~(\ref{eq:SR}), are stored in f\mbox{}iles along with the values of the incident energy and NeT corresponding to the particular MC run; 
these output f\mbox{}iles will comprise the only input when the corrections to a particular plan will be derived (planning time).

The procedure above is repeated until all options of the given proton-treatment machine have been calibrated. The variability in the material 
and in the thickness of the block is also taken into account in the current implementation (by looping over those combinations requested by 
the user~\footnote{At present, f\mbox{}ive materials are supported: brass, cerrobend, nickel, copper, and lead; this list is easily expandable. 
The block-thickness values are taken from the option properties of the machine which is conf\mbox{}igured.}). F\mbox{}inally, it is worth 
repeating that this time-consuming part of obtaining the f\mbox{}iles containing the $a_{ij}$ coef\mbox{}f\mbox{}icients (a question of a few 
hours per option) has to be performed only once, when the proton-treatment machine is conf\mbox{}igured.

At planning time, the pristine-beam f\mbox{}luence is calculated f\mbox{}irst. The beam-scattering corrections are then obtained (i.e., if 
they have been requested) after employing a number of elements developed in the course of the present section:
\begin{itemize}
\item The concept of miniblocks introduced in Section \ref{sec:Miniblocks}.
\item The reconstruction of each of the nine parameters, used in the description of the scattered protons, at a few $z$ values, on the basis 
of Eq.~(\ref{eq:SR}) from the results pertaining to the option selected in the treatment plan. The appropriate f\mbox{}ile (corresponding to the 
block material and thickness in the plan) is used as input. The f\mbox{}inal results for the various parameters are obtained via simple 2D 
interpolations in the incident energy and NeT.
\item The empirical formulae of Section \ref{sec:LatDist}.
\item A simple (linear) interpolation to obtain the corrections at all depths $z$ in the plan.
\end{itemize}
The f\mbox{}inal step involves the application of the corrections to the pristine-beam f\mbox{}luence.

The break-up of the task of determining the scattering corrections into two steps, as described in this section of the paper, enables 
the minimisation of the time consumption during the evaluation and application of these corrections in proton planning; of central 
importance, in this respect, is the concept of miniblocks.

\section{\label{sec:Results}Results}

\subsection{\label{sec:Calibration}Machine conf\mbox{}iguration}

The measurements, which are analysed in this paper, have been obtained at the Proton Therapy Centre of the National Cancer Centre (NCC), 
South Korea. The f\mbox{}irst report on the clinical commissioning and the quality assurance for proton treatment at the NCC appeared 
slightly more than one year ago, see Lee (2007).

The NCC proton-treatment machine has been manufactured by Ion Beam Applications (IBA), Louvain-la-Neuve, Belgium. Its nominal SAD is 
$2300$ mm and the distance between the f\mbox{}irst scatterer and the isocentre is $2792$ mm. The double-scattering technique currently 
supports eight options with incident energies ranging from $155$ to $230$ MeV.

\subsubsection{\label{sec:Measurements}Half-block f\mbox{}luence measurements}

\begin{figure}
\begin{center}
\includegraphics [width=15.5cm] {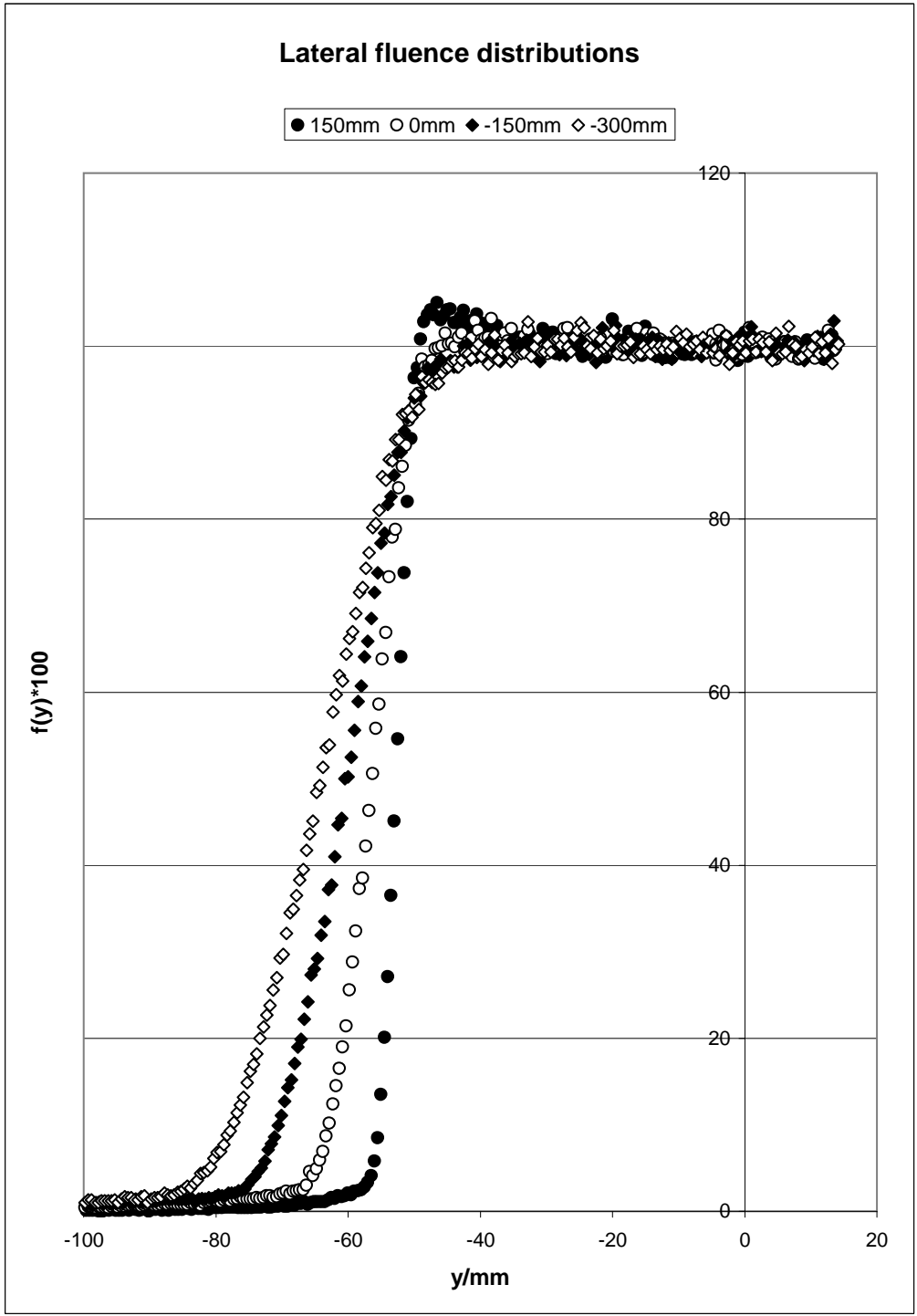}
\caption{\label{fig:HalfBlockProfiles}The lateral f\mbox{}luence distributions for one energy-NeT combination ($171.99$ MeV, $154.5$ mm) of 
option $2$ of the NCC machine.}
\end{center}
\end{figure}

All half-block measurements have been taken in air, using ${\rm IBD}=250$ mm. The lateral displacement of the $65$mm-thick brass block, used 
in these measurements, was $b=-50$ mm; the block was positioned opposite to what is shown in F\mbox{}ig.~\ref{fig:Model1}, i.e., blocking to 
beam from $-50$ mm to (theoretically) $- \infty$. To apply Eq.~(\ref{eq:FluenceOpen}) with the $t_0$ value of Eq.~(\ref{eq:t0}), the $y$ axis 
was inverted, as a result of which the $b$ value of $+50$ mm was f\mbox{}inally used in Eqs.~(\ref{eq:FluenceOpen}) and (\ref{eq:t0}). Each 
option of the double-scattering technique at the NCC comprises $15$ energy-NeT combinations; in each of these combinations, the lateral 
f\mbox{}luence distributions have been obtained at four $z$ positions, namely at $z=-300$, $-150$, $0$, and $150$ mm. One example of these 
prof\mbox{}iles is given in F\mbox{}ig.~\ref{fig:HalfBlockProfiles}. It is worth mentioning that each prof\mbox{}ile had been separately 
normalised (during the data taking) by setting the corresponding average value of the f\mbox{}luence `close' to the central beam axis to (the 
arbitrary value of) $100$~\footnote{At present, the analysis of the half-block f\mbox{}luence measurements is tedious. Unfortunately, there 
is no standard format in which these measurements, taken at the various treatment centres and machines, have to appear; in fact, there is 
complete freedom (during machine conf\mbox{}iguration) in the choice of the distance between the downstream face of the block and the isocentre, 
in the thickness and the lateral displacement of the block, and in the normalisation factors used in the output f\mbox{}iles. A more serious 
drawback is that the measurements are frequently (luckily, not in the NCC case) shifted laterally, so that the $50 \%$ of the corresponding 
maximal f\mbox{}luence (of each set, separately) be brought to $y=0$ mm; thus, an important of\mbox{}fset is irretrievably lost.}; unfortunately, 
the individual normalisation factors are not available. The `ears' of the distribution for the data set at $z=150$ mm, which are presumably 
due to block scattering, have been removed via software cuts. To avoid f\mbox{}itting the noise, f\mbox{}luence values below $10 \%$ of the 
maximal value in each data set were not processed. The block-thickness ef\mbox{}fects were removed from the data prior to processing.

Before advancing to the results, one important remark is prompt. A signif\mbox{}icant deterioration of the description of the data in the last 
of the options (option $8$) was discovered during the analysis; at present, it is not entirely clear what causes this problem. It seems that our 
beamline model leads to a systematic overestimation of the penumbra in option $8$. It has to be mentioned that the data in that option yields 
an unusually small spot size ($\sigma_s$), about $3$ times smaller than the typical values extracted in the other seven options. Option $8$ is 
the only one in which the IBA second scatterer ${\rm SS}_3$ (which is admittedly of peculiar design~\footnote{The amount of lead, used at the 
centre of the ${\rm SS}_3$, is smaller than in the cases of the ${\rm SS}_8$ and ${\rm SS}_2$ scatterers which are used in options $1$-$7$. 
Furthermore, the thickness of lexan \emph{abruptly} increases close to the rim of this scatterer. Evidently, the amount of material used in 
the ${\rm SS}_3$ cannot provide ef\mbox{}f\mbox{}icient broadening of the beam; adding material would imply smaller beam energy at nozzle exit, 
at a time when the emphasis in this option is obviously put on the high end of the energy spectrum. To be compatible with the requirement for 
f\mbox{}latness of the resulting f\mbox{}ield at isocentre, a smaller maximal f\mbox{}ield diameter ($140$ versus $220$ to $240$ mm of the other 
options) in the clinical application of option $8$ had to be imposed.}) is used. In all probability, the problems seen in option $8$ originate 
from the shape of the second scatterer, namely from the fact that the energy loss in its material is not kept constant radially (actually, it is 
`discontinuous' at $t \sim 0.9$). In order to investigate the sensitivity of our conclusions to the treatment of option $8$, we will perform the 
analysis both including in and excluding from the database the measurements of that option.

\subsubsection{\label{sec:Lambda}Extraction of parameter $\lambda$}

\begin{figure}
\begin{center}
\includegraphics [width=15.5cm] {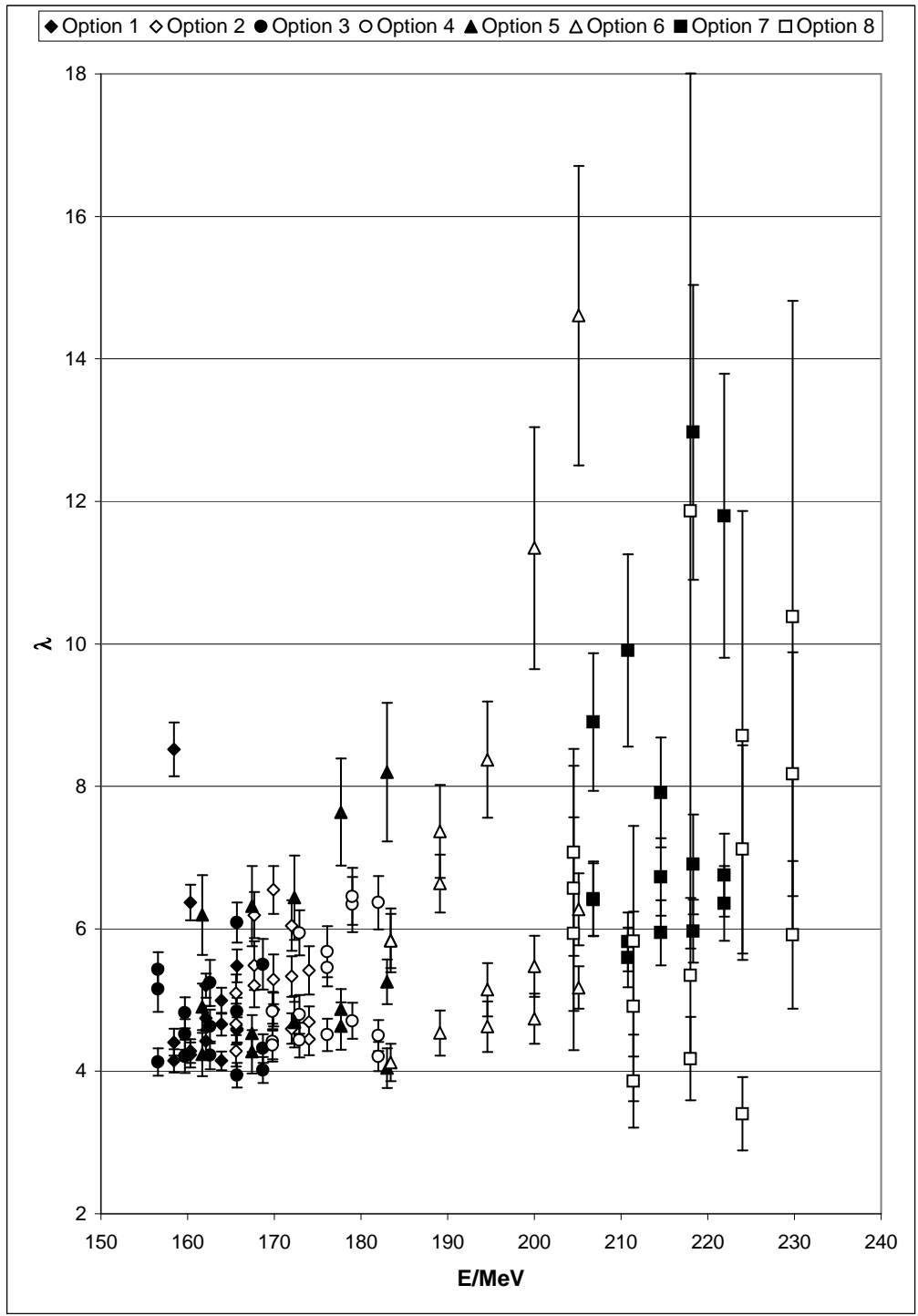}
\caption{\label{fig:Lambda}The values of the parameter $\lambda$ for all the energy-NeT combinations of all double-scattering options of the 
NCC machine.}
\end{center}
\end{figure}

The extracted values of the parameter $\lambda$ for all the energy-NeT combinations of all double-scattering options of the NCC machine are 
shown in F\mbox{}ig.~\ref{fig:Lambda}. The uncertainties in the case of option $8$ are (on average) larger than in the other options. The 
variability of the values within each option is due to the fact that, as a result of the numerous assumptions made to simplify the problem, 
$\lambda$ (which should, in principle, characterise only the second scatterer) was f\mbox{}inally turned into an ef\mbox{}fective parameter. 
To decrease the `noise' seen in F\mbox{}ig.~\ref{fig:Lambda}, the weighted average of the extracted $\lambda$ values within each option was 
f\mbox{}inally used in the ensuing MC runs for all energy-NeT combinations in that option (see Table \ref{tab:Lambda}).

\subsubsection{\label{sec:Output}Monte-Carlo runs}

\begin{figure}
\begin{center}
\includegraphics [width=15.5cm] {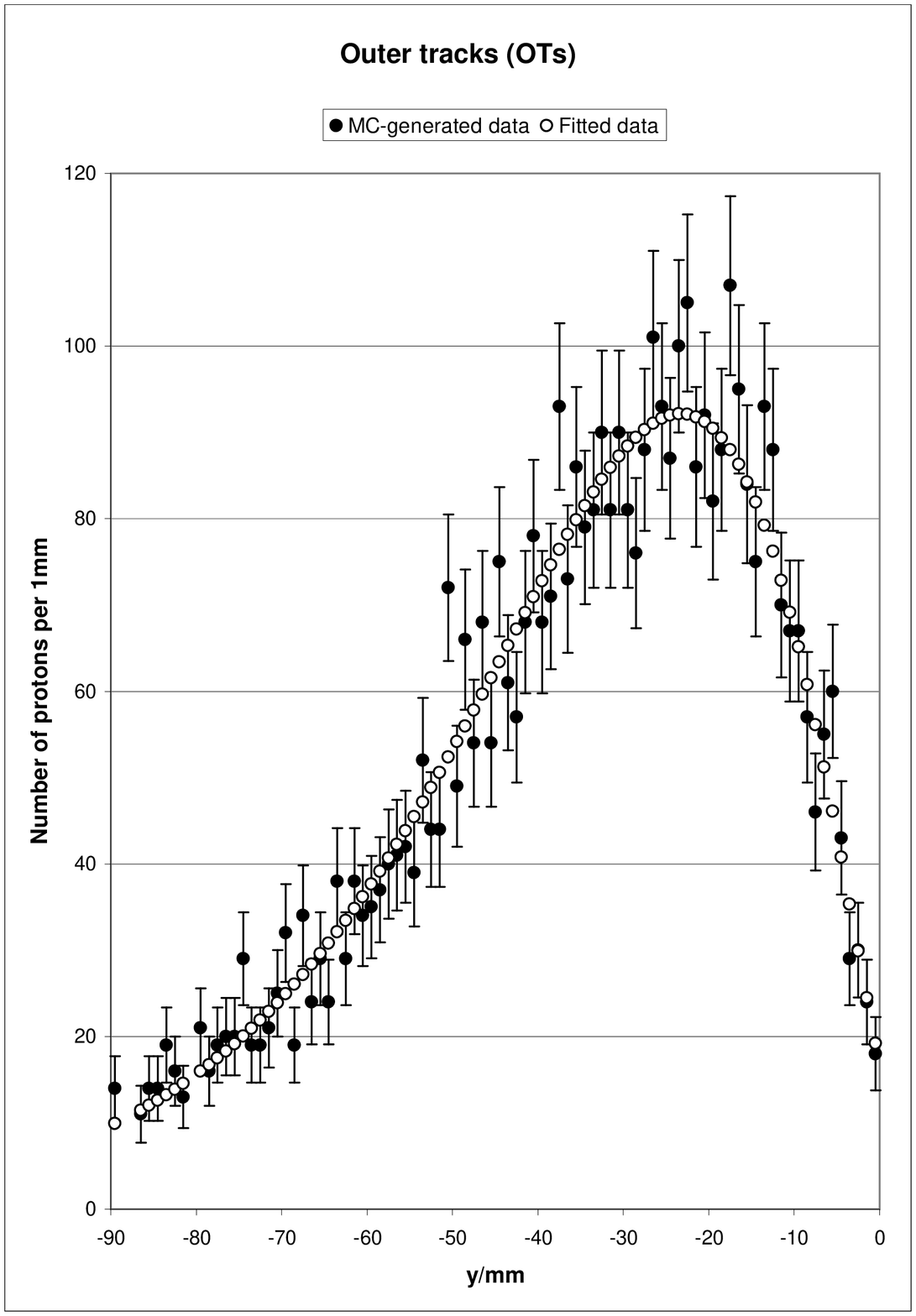}
\caption{\label{fig:Outer}A typical description of the lateral f\mbox{}luence distribution for outer tracks. Bins with fewer than $10$ entries 
are not shown.}
\end{center}
\end{figure}

\begin{figure}
\begin{center}
\includegraphics [width=15.5cm] {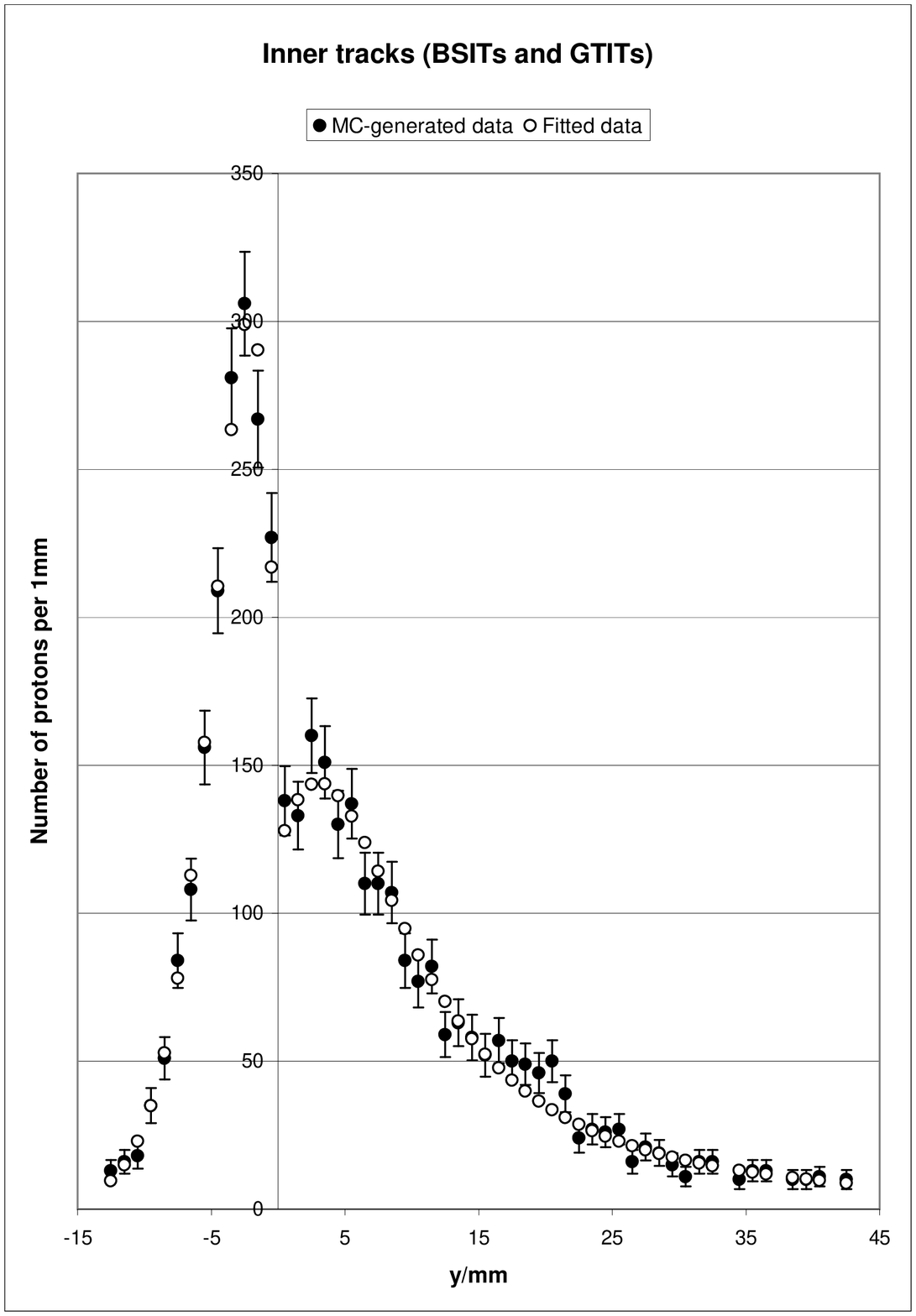}
\caption{\label{fig:Inner}A typical description of the lateral f\mbox{}luence distributions for inner tracks. Bins with fewer than $10$ entries 
are not shown.}
\end{center}
\end{figure}

$50$ million events have been generated in each energy-NeT combination, at each of three $b$ values ($20$, $40$, and $60$ mm). The lateral 
f\mbox{}luence distributions have been obtained at $17$ positions in depth, from $z=100$ to $-100$ mm. The parameters of these distributions 
for the three types of scattered protons have been extracted using the formulae of Section \ref{sec:LatDist}. In all cases (i.e., including 
option $8$), the description of the data was good; the reduced $\chi^2$ values ($\chi^2/{\rm NDF}$, NDF being the number of degrees of 
freedom in the f\mbox{}it) came out reasonably close to $1$. F\mbox{}igures \ref{fig:Outer} and \ref{fig:Inner} show typical lateral 
f\mbox{}luence distributions for outer and inner tracks, respectively.

\begin{figure}
\begin{center}
\includegraphics [width=15.5cm] {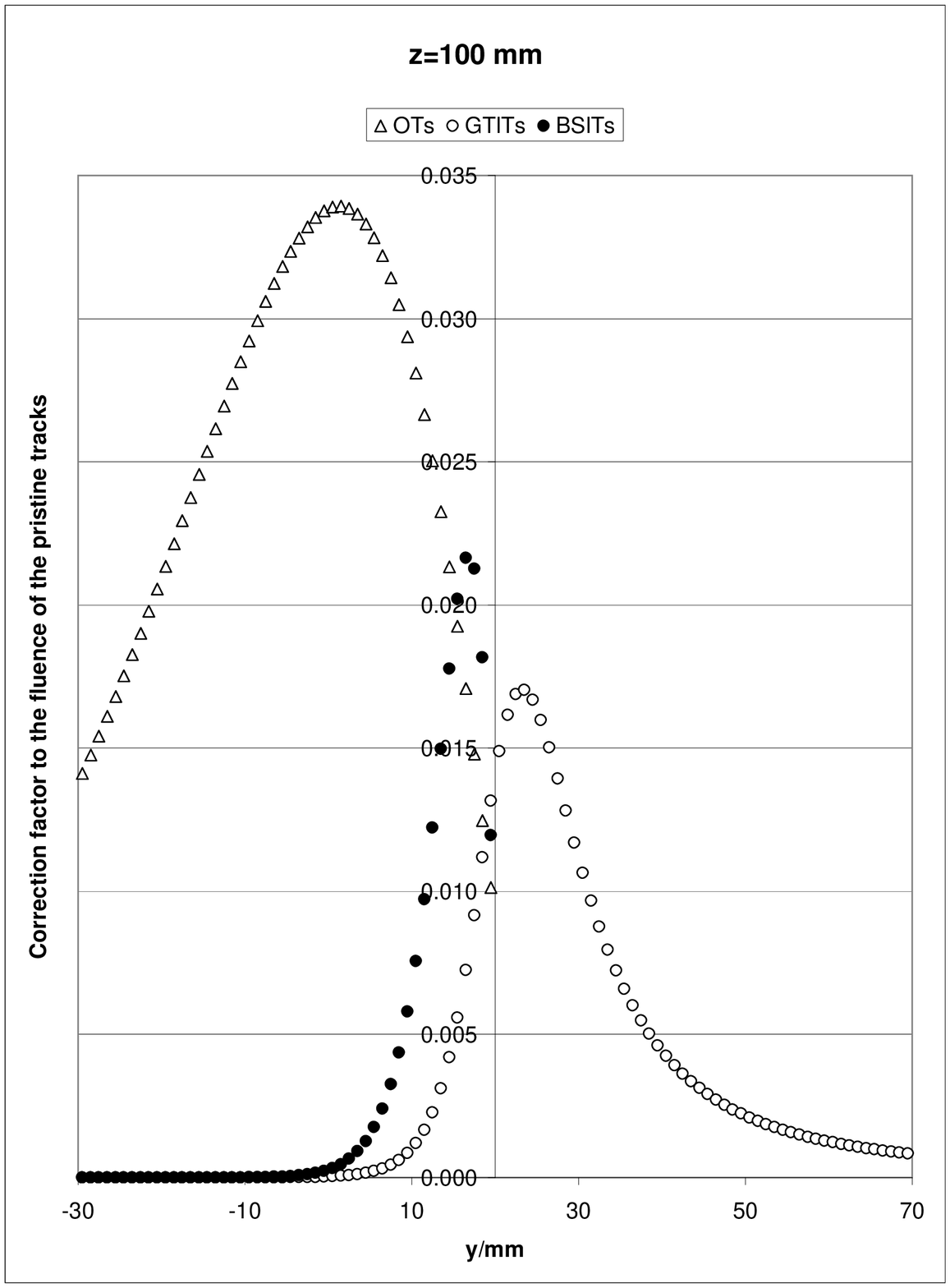}
\caption{\label{fig:CorrectionsAt100mm}One case of the scattering corrections (to be applied to the lateral f\mbox{}luence distribution of the 
pristine tracks) at $z=100$ mm. The lateral displacement of the block was $b=20$ mm.}
\end{center}
\end{figure}

\begin{figure}
\begin{center}
\includegraphics [width=15.5cm] {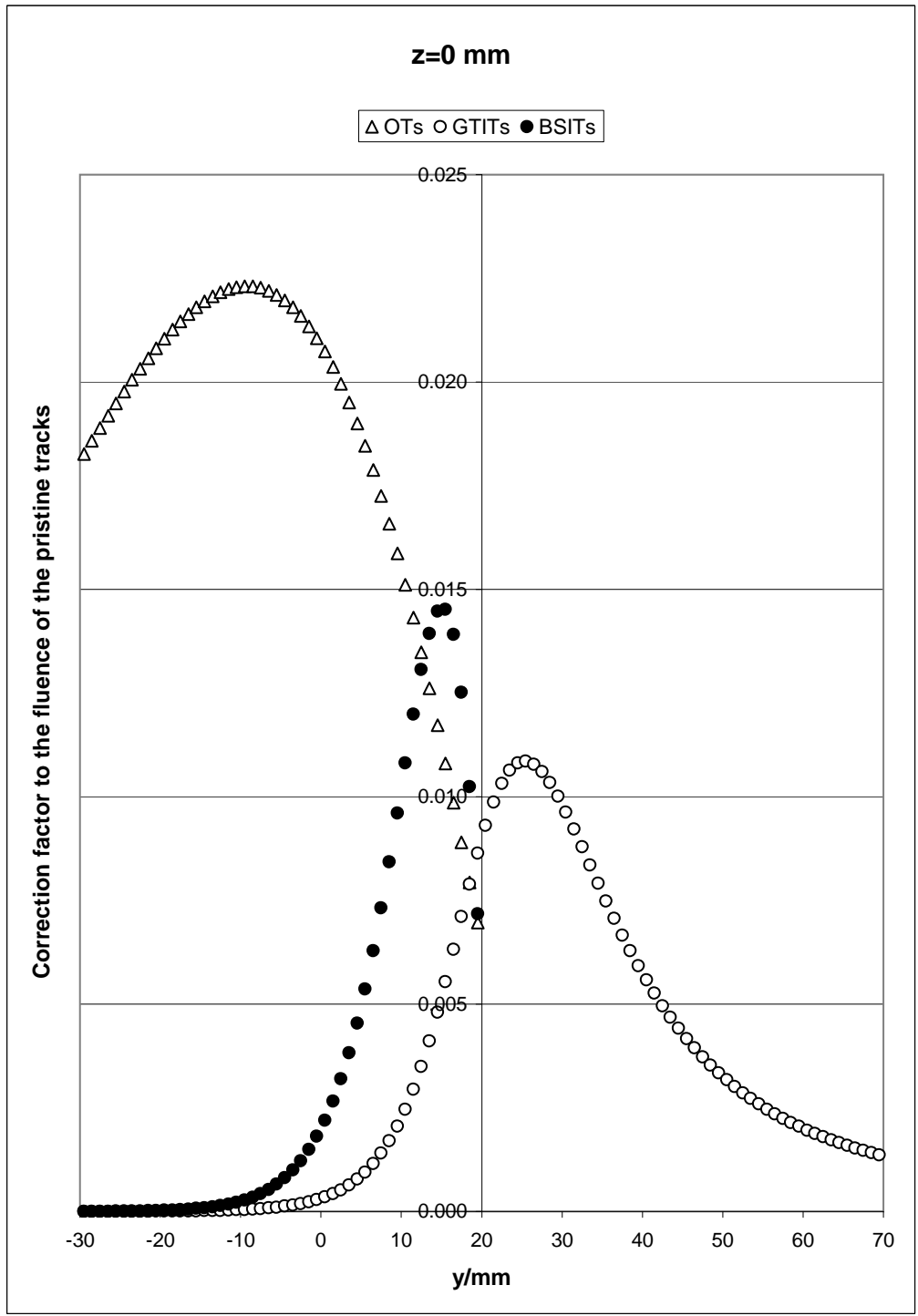}
\caption{\label{fig:CorrectionsAt0mm}One case of the scattering corrections (to be applied to the lateral f\mbox{}luence distribution of the 
pristine tracks) at $z=0$ mm (i.e., at isocentre). The lateral displacement of the block was $b=20$ mm.}
\end{center}
\end{figure}

\begin{figure}
\begin{center}
\includegraphics [width=15.5cm] {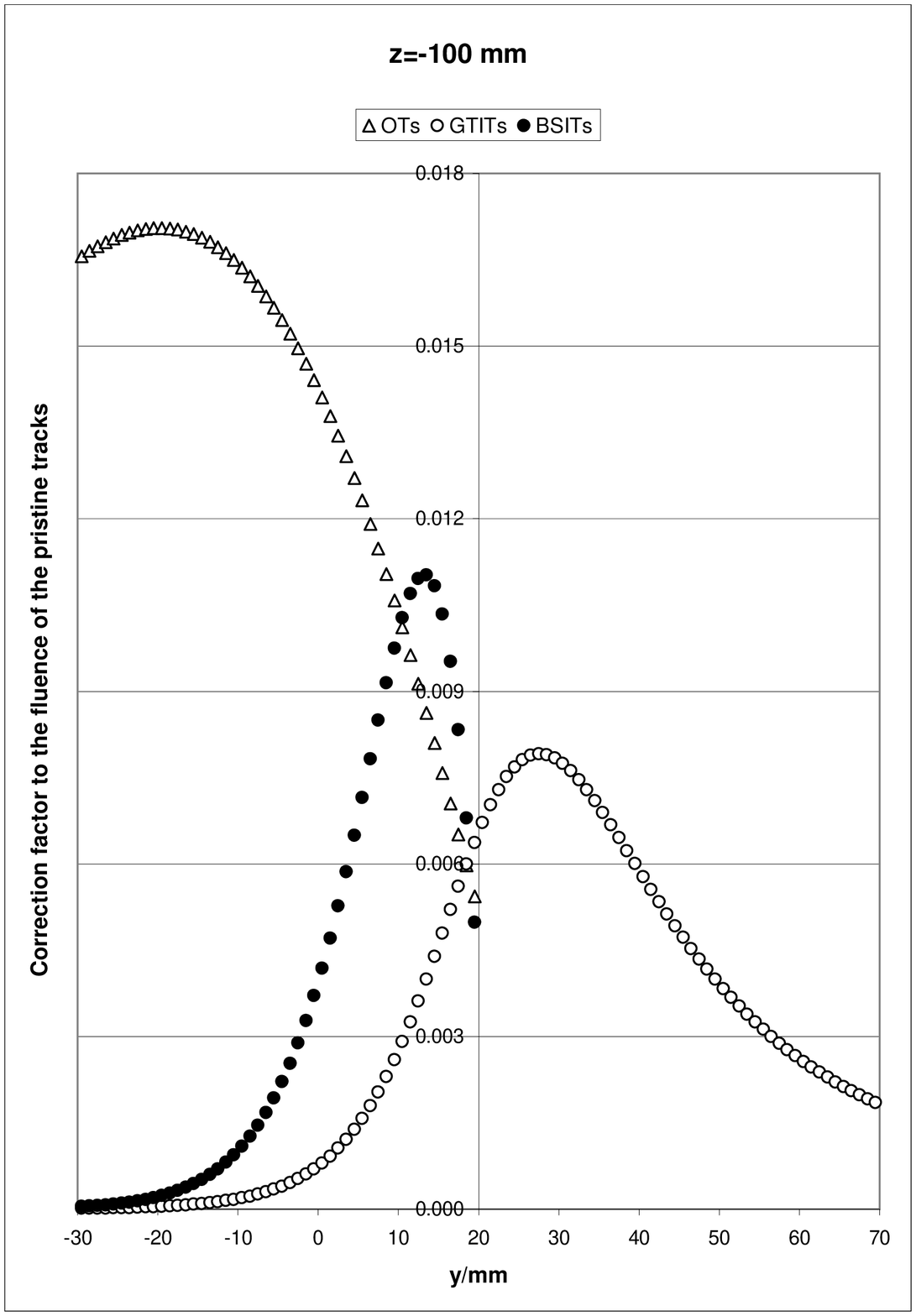}
\caption{\label{fig:CorrectionsAtM100mm}One case of the scattering corrections (to be applied to the lateral f\mbox{}luence distribution of the 
pristine tracks) at $z=-100$ mm. The lateral displacement of the block was $b=20$ mm.}
\end{center}
\end{figure}

F\mbox{}igures \ref{fig:CorrectionsAt100mm}-\ref{fig:CorrectionsAtM100mm}, obtained with a lateral block displacement of $b=20$ mm, show the 
scattering corrections (to be applied to the lateral f\mbox{}luence distributions of the pristine tracks) at three $z$ positions around the 
isocentre ($100$, $0$, and $-100$ mm, respectively). As expected, the distributions broaden when receding from the block; the mode of the 
f\mbox{}luence contribution of the OTs moves about $1$ mm away from the block extension for every $10$ mm of depth, thus indicating an `average' 
exit angle (to the bore) of about $6^0$. Concerning their magnitude, the corrections generally amount to a few percent of the corresponding 
pristine f\mbox{}luence for the typical distances involved in clinical applications.

\begin{figure}
\begin{center}
\includegraphics [width=15.5cm] {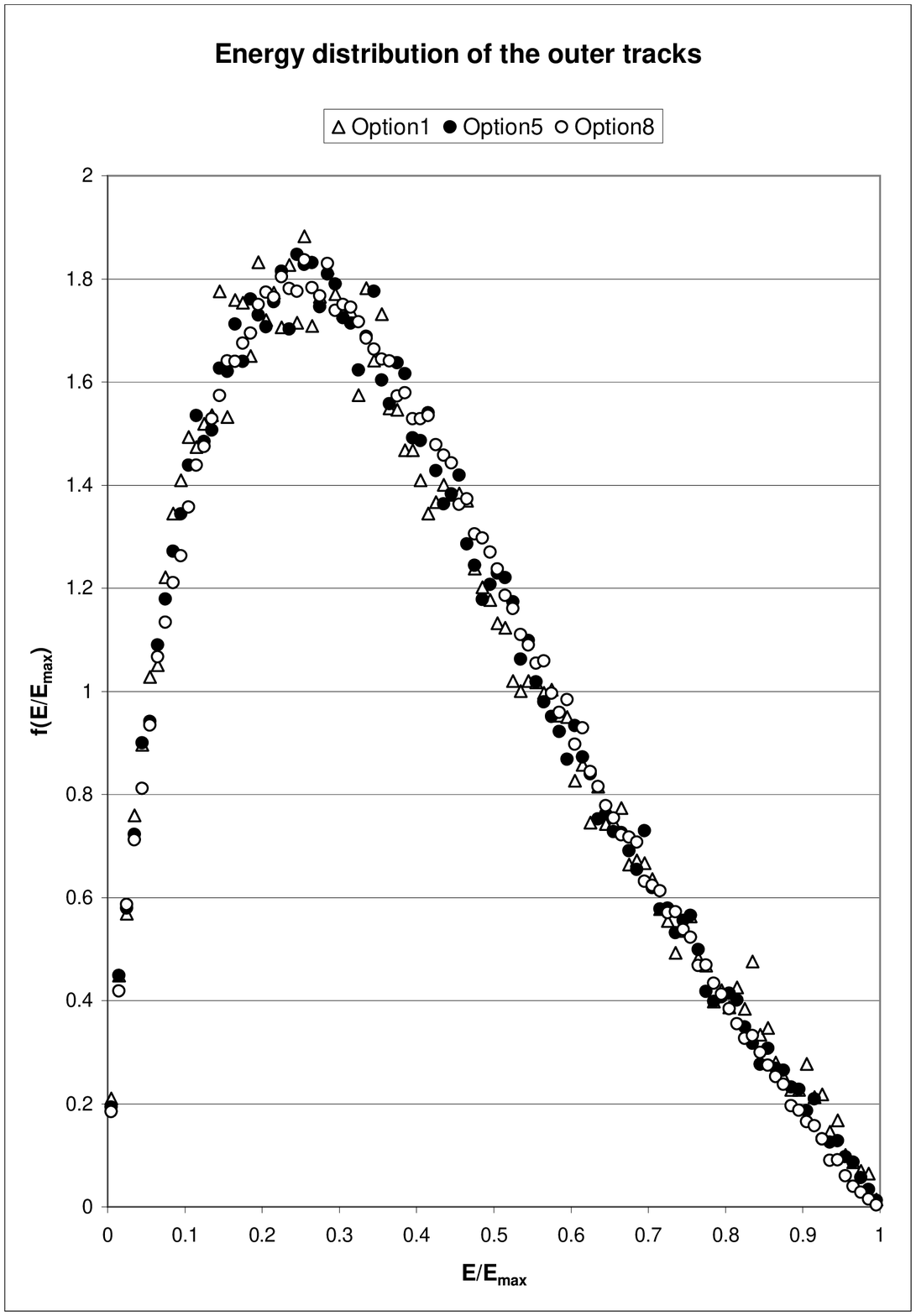}
\caption{\label{fig:EnergyOuter}The energy distributions for the outer tracks for three energy-NeT combinations of the NCC machine (see text).}
\end{center}
\end{figure}

\begin{figure}
\begin{center}
\includegraphics [width=15.5cm] {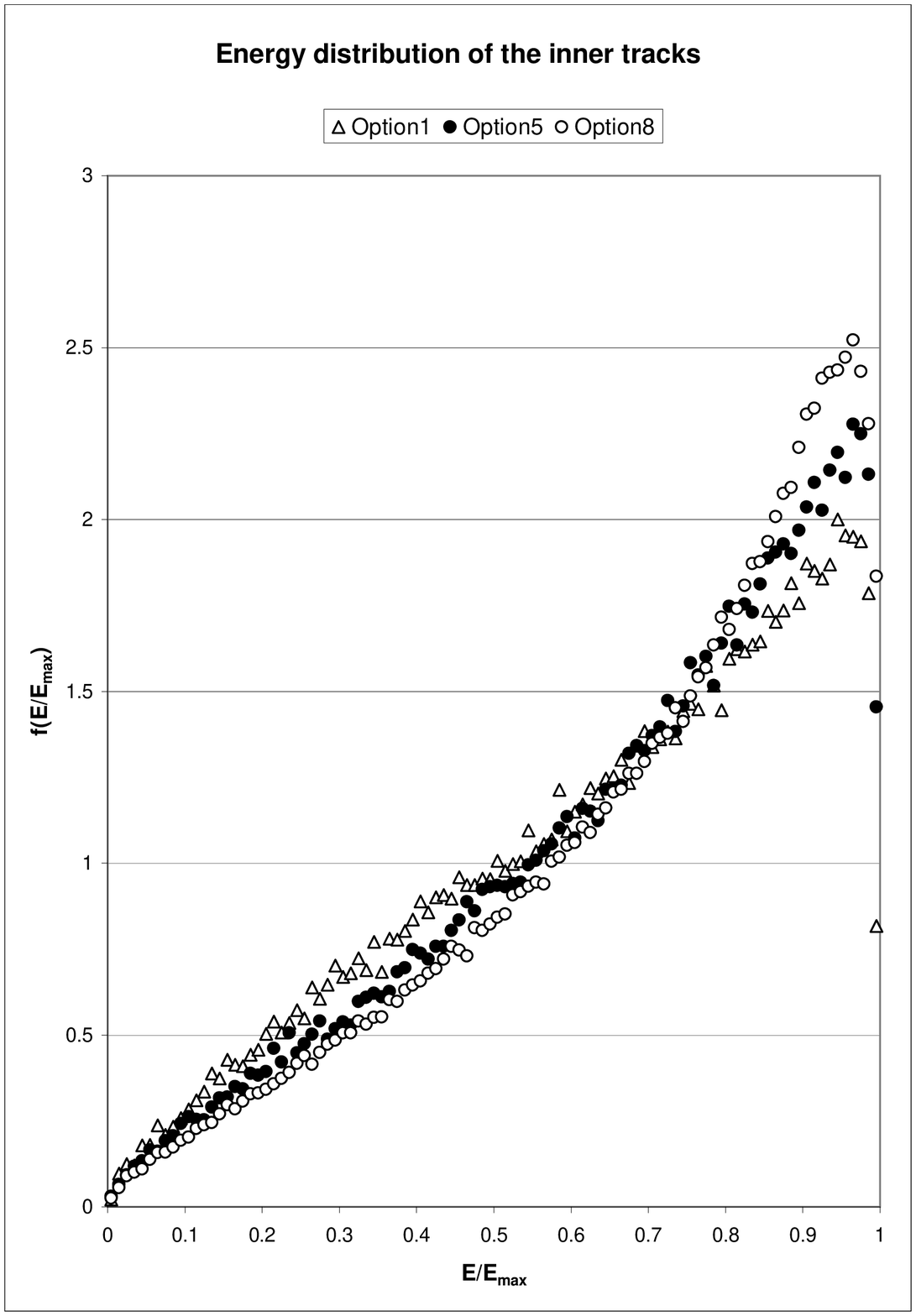}
\caption{\label{fig:EnergyInner}The energy distributions for the inner tracks for three energy-NeT combinations of the NCC machine (see text).}
\end{center}
\end{figure}

It is now time to enter the subject of the energy loss of the scattered protons. To a good approximation, one may assume that the energy 
distributions of the scattered protons depend only on the ratio $\omega=E/E_{\rm max}$, where $E$ is the energy of the scattered proton 
and $E_{\rm max}$ denotes the energy of the pristine beam (i.e., the energy at nozzle exit). In their study, Kimstrand \etal (2008) made 
the same observation. The energy distributions of the scattered protons were investigated in the case of three energy-NeT combinations 
of the NCC machine: ($158.42$ MeV, $125.3$ mm), ($182.99$ MeV, $65.6$ mm), and ($229.75$ MeV, $40.7$ mm). These combinations have been 
selected from the data sets of options $1$, $5$, and $8$, respectively, in which dif\mbox{}ferent second scatterers are employed; the 
corresponding energies $E_{\rm max}$ are: $79.5$, $150.2$, and $212.8$ MeV. The results have been obtained using a $65$-mm thick brass 
block. F\mbox{}igures \ref{fig:EnergyOuter} and \ref{fig:EnergyInner} show the energy distributions of the scattered protons as functions 
of the ratio $\omega$; instead of referring explicitly to each energy-NeT combination, we will simply use the option number (or $E_{\rm max}$) 
to identify the results. The following remarks are worth making.
\begin{itemize}
\item At low and moderate values of the exit energy $E_{\rm max}$, the energy distributions of the two types of inner tracks are similar, 
following the $f(\omega)=2 \omega$ probability distribution. At the high end of the energies used in the NCC machine, the energy distribution 
of the BSITs departs from this simplif\mbox{}ied picture, attaining a peak close to $\omega=1$; this is due to BSITs which almost `brush' the 
surface of the bore. The energy distribution of the GTITs remains unchanged. Nevertheless, to retain simplicity, we will assume that the energy 
distribution of all inner tracks follows the formula $f(\omega)=2 \omega$.
\item The energy distribution of the OTs is strongly smeared toward low $\omega$ values. The OTs lose a signif\mbox{}icant amount of their 
energy when traversing the material of the block; as shown in F\mbox{}ig.~\ref{fig:EnergyOuter}, their energy distributions peak around 
$\omega \sim 0.2$ to $0.3$. To f\mbox{}it the energy distribution of the OTs, we used the empirical formula
\[
f(\omega) \sim \sqrt{\omega} \, {\rm exp}(-\eta \, \omega) \, ,
\]
where the parameter $\eta$ turns out to be around $2$ to $2.5$; the constant of proportionality is obtained from the normalisation of the 
probability distribution.
\end{itemize}

Courant's ef\mbox{}fective-size corrections, corresponding to the three aforementioned cases, are: $0.31$, $0.87$, and $1.57$ mm. At small 
aperture sizes, the dominant contribution to the f\mbox{}luence (of the scattered protons) originates from OTs; the inner tracks dominate 
at high aperture values. The crossover is energy-dependent, ranging from about $15$ (option-$1$ result) to $47$ mm (option-$8$ result).

The absolute yields of the dif\mbox{}ferent types of the scattered protons are given in Table \ref{tab:Yields} for an aperture size (i.e., 
the diameter, assuming circular shape) of $100$ mm. We observe that the absolute yield of the OTs increases by a factor of about $2$ between 
$79.5$ and $150.2$ MeV, and by more than $4$ between $150.2$ and $212.8$ MeV. At $79.5$ MeV, the BSITs account for more than $50 \%$ of 
the total yield of the scattered protons. However, the importance of the BSITs diminishes with increasing energy, reaching the level of $1$ 
out of $4$ OTs at $212.8$ MeV. On the contrary, the yield of the GTITs f\mbox{}lattens out at about $0.75$ per OT. One might conclude that 
the BSITs seem to be more important at low energies and the OTs at high energies. The ratio of the yields of the GTITs and OTs is almost 
constant, varying only from $2/3$ to $3/4$ as the energy increases from $79.5$ to $212.8$ MeV.

\subsection{\label{sec:Verification}Verif\mbox{}ication of the method}

The verif\mbox{}ication of the method of the present paper should obviously involve the reproduction of dedicated dose measurements 
obtained in some material which, as far as the stopping power for protons is concerned, resembles human tissue, e.g., in water. The 
measurements should cover the region around the isocentre, where the tumour is usually placed, and, in order that the approach be 
validated also in the entrance region, they should extend to small distances from (the downstream face of) the block. F\mbox{}inally, 
the method must be validated for a range of depth values (associated with the energy at nozzle exit).

At present, given the lack of dedicated dose measurements, the only possibility for verif\mbox{}ication rests on re-using the calibration 
data (i.e., the half-block f\mbox{}luence measurements described in Section \ref{sec:Measurements}). We are aware of the fact that using 
the same data for conf\mbox{}iguring a system and for validating its output does not constitute an acceptable practice. However, since 
parts of the data (i.e., the areas which are obviously contaminated by the block-scattering ef\mbox{}fects) had been removed from the 
database before extracting the $\lambda$ values, this approach becomes a valid option. Luckily, as far as the validation of the scattering 
corrections to the f\mbox{}luence is concerned, our interest lies in (the reproduction of) those excluded areas; naturally, for the rest of 
the measurements (i.e., for those which \emph{were} used in the determination of the $\lambda$ values), it has to be verif\mbox{}ied that 
the quality of the description of the experimental data is not impaired by the inclusion of the block-scattering corrections.

\begin{figure}
\begin{center}
\includegraphics [width=15.5cm] {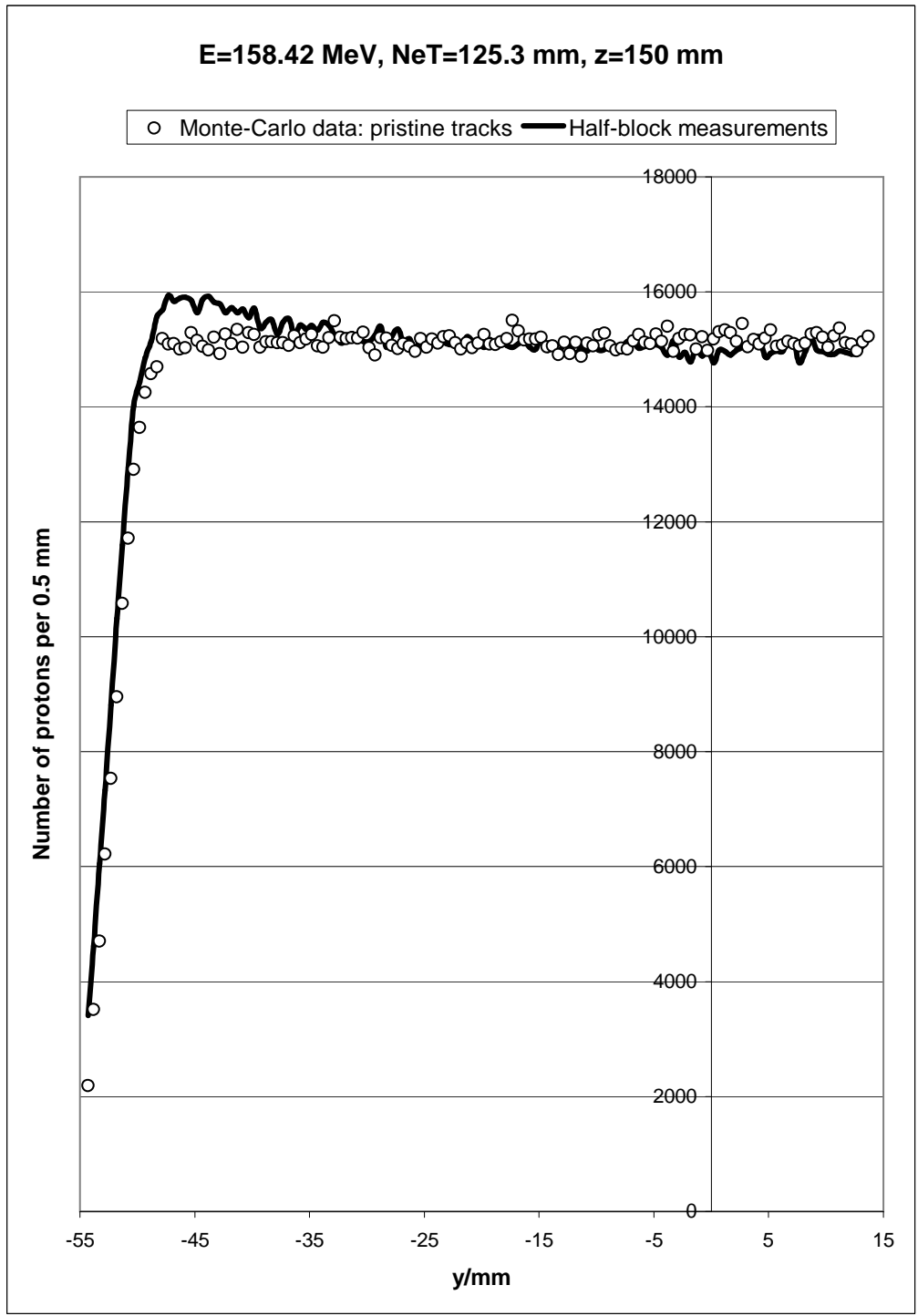}
\caption{\label{fig:Rprdctn1}The lateral f\mbox{}luence measurements (continuous line) corresponding to one energy-NeT combination of one 
option of the NCC machine, taken $100$ mm away from the downstream face of the block. The Monte-Carlo data shown correspond only to the 
pristine-beam f\mbox{}luence obtained at the same incident-energy, NeT, and $z$ values; the measurements have been scaled up by a factor 
which is equal to the ratio of the median values (of the two distributions), estimated over the f\mbox{}luence plateau.}
\end{center}
\end{figure}

\begin{figure}
\begin{center}
\includegraphics [width=15.5cm] {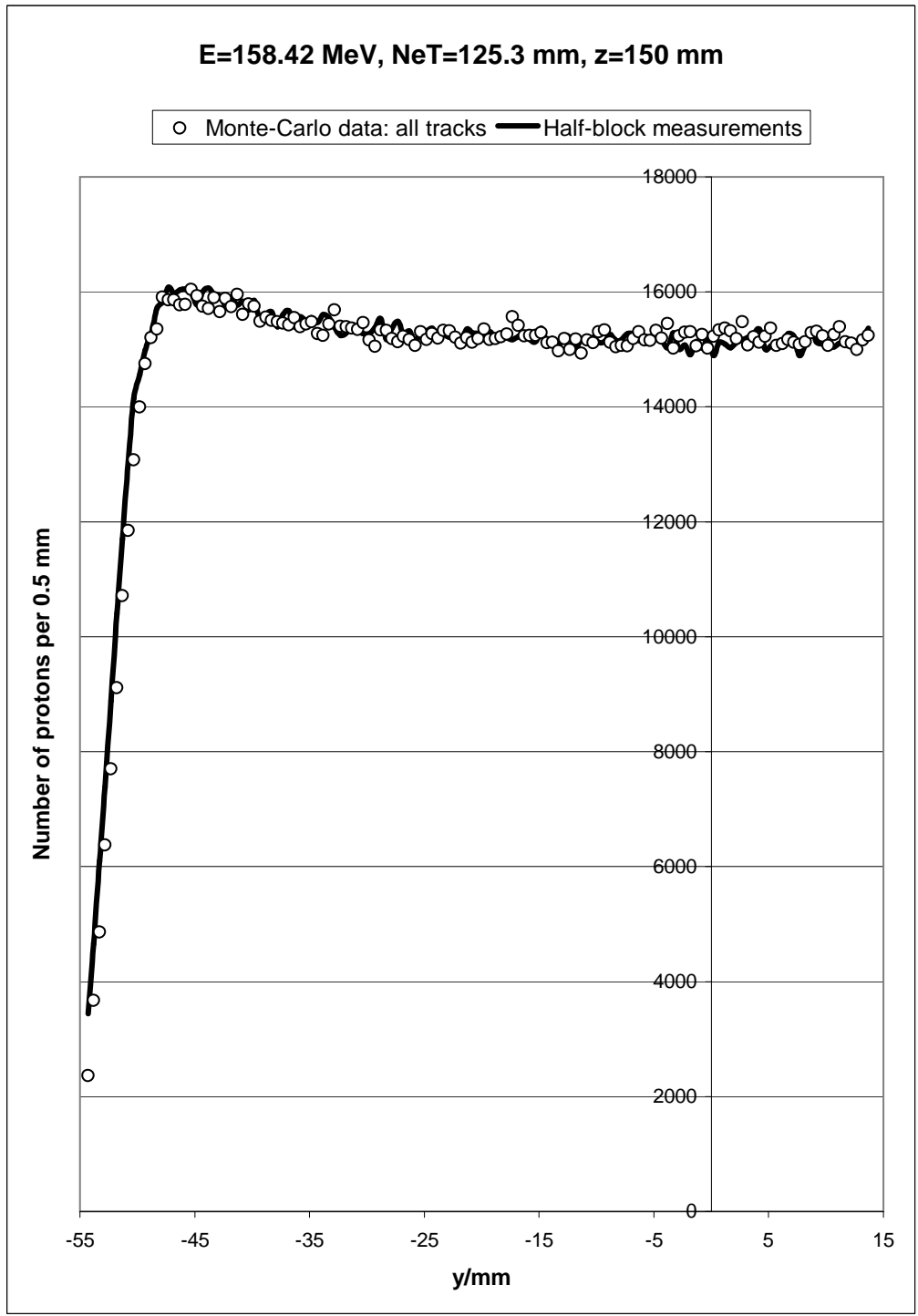}
\caption{\label{fig:Rprdctn2}The lateral f\mbox{}luence measurements (continuous line) corresponding to one energy-NeT combination of one 
option of the NCC machine, taken $100$ mm away from the downstream face of the block. The Monte-Carlo data shown correspond to the total 
(pristine-beam plus scattered-protons) f\mbox{}luence obtained at the same incident-energy, NeT, and $z$ values; the measurements have 
been scaled up by a factor which is equal to the ratio of the median values (of the two distributions), estimated over the f\mbox{}luence 
plateau.}
\end{center}
\end{figure}

A typical reproduction of the measurements is given in F\mbox{}igs.~\ref{fig:Rprdctn1} and \ref{fig:Rprdctn2}; the data correspond to the 
f\mbox{}irst energy-NeT combination of option $1$ of the NCC machine, taken at $z=150$ mm (i.e., $100$ mm away from the block). Shown in 
F\mbox{}ig.~\ref{fig:Rprdctn1} are the lateral f\mbox{}luence measurements (continuous line) along with the MC data corresponding only to 
the pristine beam; the ef\mbox{}fects of the scattered protons are added to the pristine-beam f\mbox{}luence (resulting in what will be 
called henceforth `total f\mbox{}luence') in F\mbox{}ig.~\ref{fig:Rprdctn2}. On the basis of the visual inspection of these two f\mbox{}igures, 
there is no doubt that the quality of the reproduction of the measurements in the latter case (i.e., when including the block-scattering 
ef\mbox{}fects) is superior.

We will next investigate the goodness of the reproduction of all measurements on the basis of a commonly-used statistical measure, e.g., 
of the standard $\chi^2$ function. Alternative options have been established (e.g., the $\gamma$-index approach of Low \etal (1998)), but 
have not been tried in this work. One has to bear in mind that the block-scattering contributions are larger at small distances from the 
block and in the area neighbouring its extension; at large distances, the distributions of the scattered protons broaden as a result of 
the angular divergence of the scattered beam and, less importantly, of scattering in air (an ef\mbox{}fect which has not been included 
in this paper). Evidently, the assessment of the goodness of the reproduction of the data, paying no attention to the characteristics of 
the ef\mbox{}fect in terms of the depth $z$, makes little sense.

The measurements in the area corresponding to the penumbra are very sensitive to the (input) value used for the lateral displacement of 
the block; small inaccuracy in this value af\mbox{}fects the description of the data signif\mbox{}icantly, introducing spurious 
ef\mbox{}fects in the $\chi^2$ function. This area, albeit very important in the determination of the value of the parameter $\lambda$, 
was not included in this part of the analysis.

The resulting $\chi^2$ values are given in Table \ref{tab:ChiSquare}, separately for the four $z$ positions at which the 
measurements have been obtained; it is evident that, for all depth values, the quality of the reproduction of the measurements when 
including the block-scattering ef\mbox{}fects is superior to the case that only the pristine-beam contribution is considered. The 
importance of the inclusion of these ef\mbox{}fects decreases with increasing distance to the block. The improvement for $z=150$ mm 
when including the block-scattering ef\mbox{}fects is impressive. Judging from the $\chi^2$ values for the given degrees of freedom, 
there can be no doubt that the overall reproduction of the data is satisfactory. Last but not least, despite the fact that the description 
of the option-$8$ measurements is debatable, our conclusions do not depend on the treatment of the data in that option (i.e., inclusion in 
or exclusion from the database).

Out of the available $120$ lateral f\mbox{}luence prof\mbox{}iles, corresponding to $z=150$ mm, only f\mbox{}ive prof\mbox{}iles did not 
show improvement after the scattering ef\mbox{}fects were included. In two of these prof\mbox{}iles, namely the energy-NeT combinations 
of ($176.08$ MeV, $102.2$ mm) and ($179.00$ MeV, $103.3$ mm), the scattering contributions are present in the measurements, yet at 
dif\mbox{}ferent amounts compared to the MC-generated data; additionally, a hard-to-explain slope (i.e., of dif\mbox{}ferent sign 
to what is expected after including the scattering contributions) is clearly seen in these measurements around $y=0$ mm. On the other 
hand, no scattering ef\mbox{}fects (`ears') are discernible in the ($167.42$ MeV, $160.5$ mm), ($206.77$ MeV, $75.2$ mm), and ($204.47$ MeV, 
$203.2$ mm) combinations at $z=150$ mm. It has to be stressed that these f\mbox{}ive data sets are surrounded by a multitude of measurements 
showing an impressive agreement between the experimentally-obtained and the MC-generated total-f\mbox{}luence distributions; due to this 
reason, we rather consider the absence of improvement (that is, after the scattering contributions are included) in the description of the 
data in these f\mbox{}ive prof\mbox{}iles as indicative of experimental problems.

\begin{figure}
\begin{center}
\includegraphics [width=15.5cm] {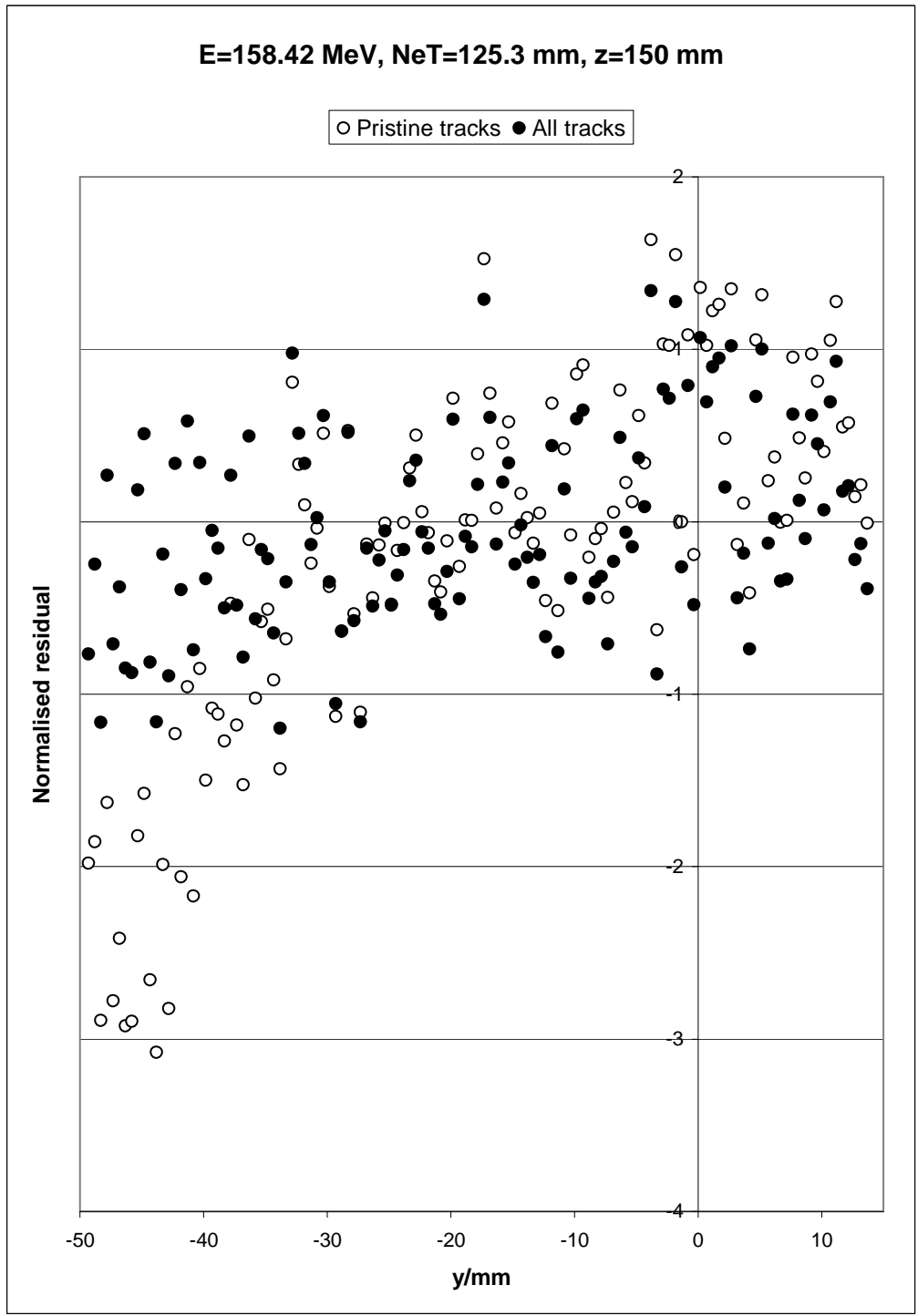}
\caption{\label{fig:NormRes}An alternative way of displaying the contents of F\mbox{}igs.~\ref{fig:Rprdctn1} and \ref{fig:Rprdctn2}; shown 
in this f\mbox{}igure are the normalised residuals, plotted versus the lateral distance $y$. Evidently, the pristine-beam contribution 
underestimates the f\mbox{}luence by about $1$ to $3$ standard deviations for $-50$ mm $ < y < -40$ mm. After the inclusion of the 
block-scattering ef\mbox{}fects, the residuals nicely cluster around $0$.}
\end{center}
\end{figure}

An alternative way of displaying the content of F\mbox{}igs.~\ref{fig:Rprdctn1} and \ref{fig:Rprdctn2} is given in 
F\mbox{}ig.~\ref{fig:NormRes}; instead of showing separately the measurements and the MC data, shown in F\mbox{}ig.~\ref{fig:NormRes} 
are the normalised residuals, def\mbox{}ined as
\[
Z_i=\frac{v_i^{\rm MC}-v_i^{\rm exp}}{\sqrt{(\delta v_i^{\rm MC})^2+(\delta v_i^{\rm exp})^2}} \, ,
\]
where $v_i^{\rm exp}$ denotes the $i$-th measurement and $v_i^{\rm MC}$ the corresponding MC-obtained f\mbox{}luence; $\delta v_i^{\rm exp}$ 
and $\delta v_i^{\rm MC}$ represent their uncertainties. The advantage of such a plot is evident as direct information on the reproduction 
may be obtained faster than from F\mbox{}igs.~\ref{fig:Rprdctn1} and \ref{fig:Rprdctn2}; for instance, not only can one immediately become 
aware of the failure of the pristine-beam data close to the borders of the block, but also of the severity of this failure. Evidently, the 
pristine-beam contribution underestimates the f\mbox{}luence by about $1$ to $3$ standard deviations for $-50$ mm $ < y < -40$ mm. It is 
also interesting to note that, after the scattering ef\mbox{}fects have been included, the normalised residuals $Z_i$ show 
signif\mbox{}icantly smaller dependence on the lateral distance $y$ (ideally, no dependence should be seen).

\subsection{\label{sec:Planning}An example of the application of the corrections}

The aim of the present paper was to provide the systematic description of a method to be used in the determination and application 
of the corrections which are due to the presence of BL/BSDs in proton planning. Despite the fact that no emphasis was meant to be put on 
a clinical investigation, one simple example (of the application of the corrections) may nevertheless be called for.

\begin{figure}
\begin{center}
\includegraphics [width=15.5cm] {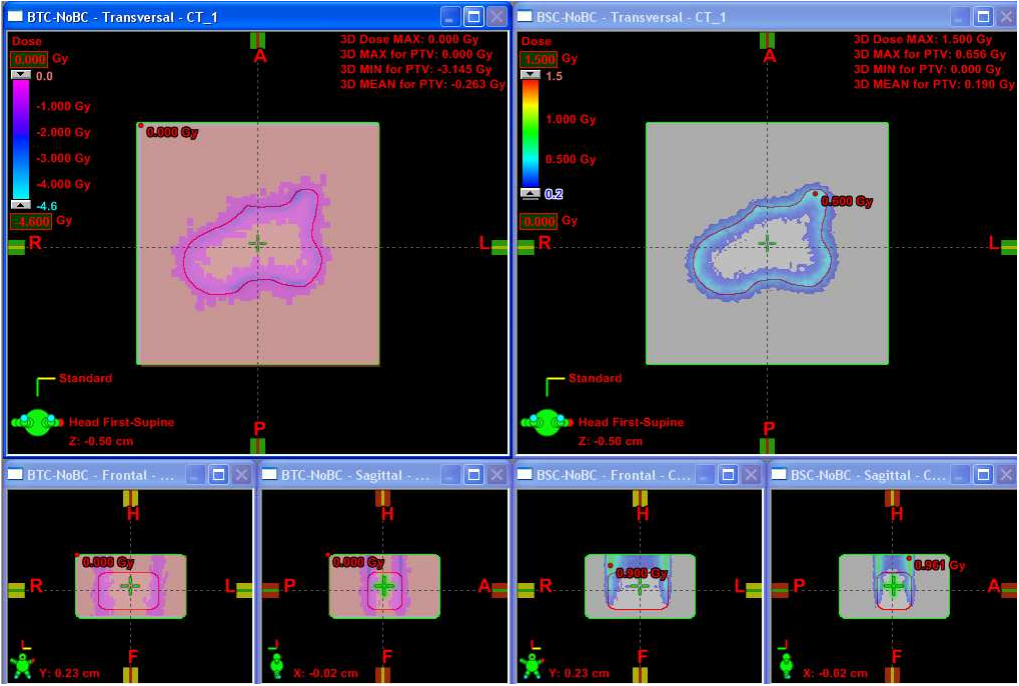}
\caption{\label{fig:Eclipse1}The dose contributions corresponding to the block-thickness (on the left) and to the block-scattering (on 
the right) corrections for the simple water phantom of Section \ref{sec:Planning}.}
\end{center}
\end{figure}

\begin{figure}
\begin{center}
\includegraphics [width=15.5cm] {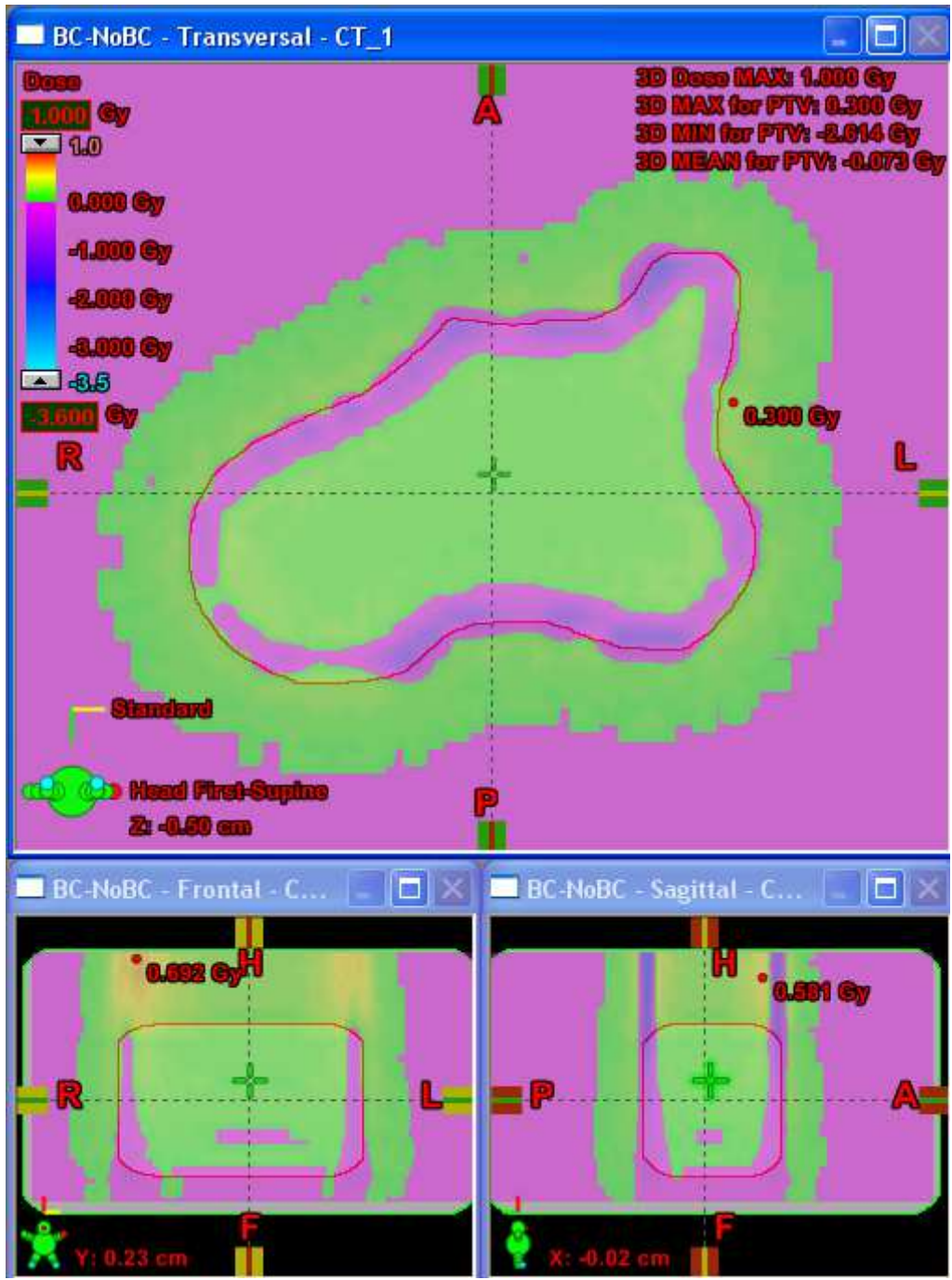}
\caption{\label{fig:Eclipse2}The dose contributions corresponding to both block-relating corrections for the simple water phantom of 
Section \ref{sec:Planning}.}
\end{center}
\end{figure}

To this end, the current version of the Treatment Planning System Eclipse$^{\rm (TM)}$ (Varian Medical Systems Inc., Palo Alto, California) 
was extensively modif\mbox{}ied to include the derivation (in beam conf\mbox{}iguration) and the application (in planning) of both block-relating 
corrections; the application of each of the two corrections may be requested separately in the user interface. (A review article, providing 
the details of the dose evaluation in proton therapy, as well as its implementation in Eclipse, will appear soon, see Ulmer \etal (2009).) A 
simple rectangular water phantom was created, within which a planning treatment volume (PTV) of $90 \, {\rm cm}^3$ was arbitrarily outlined. 
Four one-f\mbox{}ield treatment plans were subsequently created as follows:
\begin{itemize}
\item[] a) a plan without block-relating corrections,
\item[] b) a plan with block-thickness corrections,
\item[] c) a plan with block-scattering corrections, and
\item[] d) a plan with both block-relating corrections~\footnote{To void double counting in the case that both block-relating 
corrections are requested, the block-thickness corrections are f\mbox{}irst applied to the pristine-beam f\mbox{}luence. The 
block-scattering corrections, which already contain their corresponding block-thickness ef\mbox{}fects, are subsequently added 
to the `thickness-corrected' pristine-beam f\mbox{}luence.}.
\end{itemize}
In each of these plans, a brass block was inserted and f\mbox{}itted to the cross section of the PTV. Subsequently, a dose of $100$ Gy 
was delivered to the PTV, using the double-scattering technique of the NCC machine. The block-relating corrections were estimated on 
the basis of $N=64$; in fact, the results are practically insensitive to the value of $N$, for $N \geq 16$. F\mbox{}inally, the resulting 
dose maps were compared; the dose dif\mbox{}ferences of plans (b), (c), and (d) to plan (a) were estimated and compared 
(F\mbox{}igs.~\ref{fig:Eclipse1} and \ref{fig:Eclipse2}), leading to the following conclusions.
\begin{itemize}
\item As expected, the application of the block-thickness corrections results in lower dose values. This is due to the fact that part 
of the incident f\mbox{}lux is blocked as a result of the nonzero thickness of the block.
\item As expected, the application of the block-scattering corrections leads to higher dose values. This is because some protons, 
which would otherwise fail to contribute to the f\mbox{}luence (as they impinge upon the block), scatter of\mbox{}f the material 
of the block and `re-emerge' at positions in the bore or on the downstream face of the block.
\item As far as the delivered dose is concerned, the ef\mbox{}fects of the block thickness and block scattering `compete' one 
another. In the water phantom used in this section, the block-thickness corrections dominate.
\item The presence of the block in the plan of the water phantom used in this section induces ef\mbox{}fects which amount to a few 
percent of the prescribed dose (see F\mbox{}ig.~\ref{fig:Eclipse2}). The largest ef\mbox{}fects appear in the area neighbouring the 
border of the block. It is also worth noticing the characteristic contributions of the scattered protons in the entrance region in 
the frontal and sagittal views in F\mbox{}ig.~\ref{fig:Eclipse2}; the largest part of the dose in the entrance region corresponds to 
the low-energy component of the scattered beam. As the \emph{local} dose, delivered in the entrance region, is signif\mbox{}icantly smaller 
than the corresponding value delivered to the target (which, in fact, is an agrument in favour of the use of protons in radiation therapy), 
the corrections which one has to apply to it, though representing a small fraction of the \emph{prescribed} dose, are sizable.
\end{itemize}

\section{\label{sec:Conclusions}Conclusions}

The present work deals with corrections which are due to the presence of beam-limiting and beam-shaping devices in proton planning. 
The application of these corrections is greatly facilitated by decomposing the ef\mbox{}fects of two-dimensional objects into 
one-dimensional, easily-calculable contributions (miniblocks).

In the derivation of the thickness corrections, we follow the strategy of Slopsema and Kooy (2006). Given the time restrictions during 
the planning, the derivation of the scattering corrections necessitates the introduction of a two-step approach. The f\mbox{}irst step 
occurs at the beam-conf\mbox{}iguration phase. At f\mbox{}irst, the value of the only parameter of our model ($\lambda$) is extracted 
from the half-block f\mbox{}luence measurements. A number of Monte-Carlo runs follow, the output of which consists of the parameters 
pertaining to convenient parameterisations of the f\mbox{}luence contributions of the scattered protons. These runs take account of the 
variability in the block material and thickness, incident energy, and nozzle-equivalent thickness in all the options for which a 
proton-treatment machine is conf\mbox{}igured. To enable the easy use of the MC results, the output is put in the form of expansion 
parameters in two geometrical quantities which are involved in the description of the scattering ef\mbox{}fects. The scattering corrections for 
all the blocks in a particular plan are determined from the results, obtained at beam-conf\mbox{}iguration phase, via simple interpolations.

The verif\mbox{}ication of the method should involve the reproduction of dedicated dose measurements. At present, given the lack of 
such measurements, the only possibility for verif\mbox{}ication rested on re-using the half-block f\mbox{}luence measurements, 
formerly analysed to extract the $\lambda$ value; this is a valid option because parts of the input data had been removed from the 
database to suppress the (present in the measurements) block-scattering contributions. We investigated the goodness of the reproduction 
of the measurements on the basis of the $\chi^2$ function and concluded that the inclusion of the scattering ef\mbox{}fects 
leads to substantial improvement.

The method presented in this paper was applied to one plan involving a simple water phantom; the dif\mbox{}ferent contributions from 
the two block-relating ef\mbox{}fects have been separately presented and compared. These ef\mbox{}fects amount to a few percent of 
the prescribed dose and are signif\mbox{}icant in the entrance region and in the area neighbouring the border of the block.

\begin{ack}
The author acknowledges helpful discussions with Barbara Schaf\mbox{}fner concerning the optimal implementation of the method in Eclipse. 
Barbara also modelled and implemented in Eclipse the important (in the entrance region) dose contributions of the low-energy scattered 
protons (i.e., those with energies below the lowest value used in the particular plan to `spread out' the Bragg peak). The author is 
grateful to Se Byeong Lee for providing the original NCC half-block f\mbox{}luence measurements, as well as important information on the 
data taking.
\end{ack}

\References
\item[] Burge E J and Smith D A 1962 Theoretical Study Of Slit Scattering {\it Rev. Sci. Instrum.} {\bf 33} 1371
\item[] Courant E D 1951 Multiple Scattering Corrections for Collimating Slits {\it Rev. Sci. Instrum.} {\bf 22} 1003
\item[] Geant4, Physics Reference Manual, version: geant4 9.1 (14 December, 2007); available from the internet address 
http://geant4.web.cern.ch/geant4/support/userdocuments.shtml
\item[] Gottschalk B, Koehler A M, Schneider R J, Sisterson J M and Wagner M S 1993 Multiple Coulomb scattering of 160 MeV protons 
{\it Nucl. Instr. Meth.} {\bf B74} 467
\item[] Highland V L 1975 Some practical remarks on multiple scattering {\it Nucl. Instr. Meth.} {\bf 129} 497
\item[] ICRU Report 1987 Use of Computers in External Beam Radiotherapy Procedures with High-Energy Photons and Electrons {\bf 42} 1
\item[] Kimstrand P, Traneus E, Ahnesj{\"o} A and Tilly N 2008 Parametrization and application of scatter kernels for modelling 
scanned proton beam collimator scatter dose {\it Phys. Med. Biol.} {\bf 53} 3405
\item[] Lee S B 2007 Clinical Commissioning and Quality Assurance of Proton beam in NCC, Korea {\it PTCOG 46}; available from the 
internet address http://ptcog.web.psi.ch/PTCOG46
\item[] Low D A, Harms W B, Mutic S and Purdy J A 1998 A technique for the quantitative evaluation of dose distributions {\it Med. Phys.} 
{\bf 25} 656
\item[] Lynch G R and Dahl O I 1991 Approximations to multiple Coulomb scattering {\it Nucl. Instr. Meth.} {\bf B58} 6
\item[] Minuit2 Minimization Package, 5.20/00; available from the internet address 
http://project-mathlibs.web.cern.ch/project-mathlibs/sw/Minuit2/html/index.html
\item[] Paganetti H 1998 Monte Carlo method to study the proton f\mbox{}luence for treatment planning {\it Med. Phys.} {\bf 25} 2370
\item[] Slopsema R L and Kooy H M 2006 Incorporation of the aperture thickness in proton pencil-beam dose calculations {\it Phys. Med. Biol.} 
{\bf 51} 5441
\item[] Takada Y 2002 Optimum solution of dual-ring double-scattering system for an incident beam with given phase space for proton beam 
spreading {\it Nucl. Instr. Meth.} {\bf A485} 255
\item[] Ulmer W, Matsinos E and Kaissl W 2009 Theoretical methods for the calculation of Bragg curves and 3D distributions of proton beams; 
to appear in {\it Radiation Physics and Chemistry}
\item[] van Luijk P, van 't Veld A A, Zelle H D and Schippers J M 2001 Collimator scatter and 2D dosimetry in small proton beams 
{\it Phys. Med. Biol.} {\bf 46} 653
\item[] Yao W-M \etal 2006 The Review of Particle Physics {\it J. Phys.} {\bf G33} 1; available from the internet address http://pdg.lbl.gov/
\endrefs

\newpage
\vspace{0.5cm}
\begin{table}
\caption{\label{tab:Lambda}The weighted averages of the extracted $\lambda$ values (and their statistical uncertainties) for the eight options 
of the NCC machine. The uncertainties are shown only for the sake of completeness; they have not been taken into account in the results of 
Section \ref{sec:Output}.}
\vspace{0.2cm}
\begin{center}
\begin{tabular}{|c|c|}
\hline
Option number & $\lambda (\delta\lambda)$ \\
\hline
$1$ & $4.667 \, (0.046)$ \\
$2$ & $5.020 \, (0.070)$ \\
$3$ & $4.549 \, (0.057)$ \\
$4$ & $4.857 \, (0.069)$ \\
$5$ & $4.800 \, (0.090)$ \\
$6$ & $5.26 \, (0.11)$ \\
$7$ & $6.42 \, (0.15)$ \\
$8$ & $4.65 \, (0.27)$ \\
\hline
\end{tabular}
\end{center}
\end{table}
\vspace{0.5cm}
\begin{table}
\caption{\label{tab:Yields}The absolute yields (numbers of particles) of the dif\mbox{}ferent types of the scattered protons for an aperture 
size of $100$ mm, for three energy-NeT combinations of the NCC machine (see text); a $65$-mm thick brass block has been used.}
\vspace{0.2cm}
\begin{center}
\begin{tabular}{|c|c|c|c|}
\hline
$E_{\rm max}$ (MeV) & OTs & BSITs & GTITs \\
\hline
$79.5$ & $35696$ & $64130$ & $23799$ \\
$150.2$ & $72015$ & $42128$ & $54383$ \\
$212.8$ & $319806$ & $83808$ & $242229$ \\
\hline
\end{tabular}
\end{center}
\end{table}
\vspace{0.5cm}
\begin{table}
\caption{\label{tab:ChiSquare}The $\chi^2$ values, corresponding to the reproduction of the half-block f\mbox{}luence measurements, separately 
for the four $z$ positions at which the data have been obtained; evidently, the quality of the reproduction of the measurements when 
including the block-scattering ef\mbox{}fects is superior to the case that only the pristine-beam contribution is taken into account. NDF denotes 
the number of degrees of freedom. The lower part of the table contains the results after excluding the measurements of option $8$.}
\vspace{0.2cm}
\begin{center}
\begin{tabular}{|c|c|c|c|}
\hline
$z$ (mm) & $\chi^2$ Pristine & $\chi^2$ Total & NDF \\
\hline
$150$ & $22710.25$ & $8551.34$ & $14511$ \\
$0$ & $15384.09$ & $11826.71$ & $14210$ \\
$-150$ & $13229.26$ & $12255.67$ & $13841$ \\
$-300$ & $12949.22$ & $12573.03$ & $13628$ \\
\hline
$150$ & $16024.44$ & $5761.21$ & $12704$ \\
$0$ & $12001.81$ & $9761.32$ & $12404$ \\
$-150$ & $10751.43$ & $10277.58$ & $12057$ \\
$-300$ & $10658.78$ & $10531.18$ & $11843$ \\
\hline
\end{tabular}
\end{center}
\end{table}
\clearpage
\clearpage
\clearpage
\clearpage
\clearpage
\clearpage
\clearpage
\clearpage
\clearpage
\clearpage
\clearpage
\clearpage
\clearpage
\clearpage
\clearpage
\clearpage
\clearpage
\clearpage
\clearpage
\clearpage
\clearpage
\clearpage
\clearpage

\end{document}